\documentclass[a4paper,12pt]{article}
\pdfoutput=1 
\usepackage{jheppub} 
\usepackage{color}

\newcommand{\be}{\begin{equation}}
\newcommand{\ee}{\end{equation}}
\newcommand{\beal}{\begin{aligned}}
\newcommand{\eeal}{\end{aligned}}
\newcommand\bea {\begin{eqnarray}}
\newcommand\eea {\end{eqnarray}}
\newcommand{\bec}{\begin{cases}}
\newcommand{\eec}{\end{cases}}
\newcommand{\bei}{\begin{itemize}}
\newcommand{\eei}{\end{itemize}}
\newcommand{\bee}{\begin{enumerate}}
\newcommand{\eee}{\end{enumerate}}

\newcommand{\sech}{\rm{sech}}

\newcommand{\bear}{\begin{array}}
\newcommand{\enar}{\end{array}}

\newcommand{\vthermo}{\mathcal{V}}

\def\be{\begin{equation}}
\def\ee{\end{equation}}
\def\bea{\begin{eqnarray}}
\def\eea{\end{eqnarray}}

\def\dphi{\dot{\phi}}

\def\da{\dot{a}}
\def\phif{ \phi_f }

\def\ddvf{ W^{\prime\prime}_f }
\def\vcal{\mathcal V}

\title{Black Hole Thermodynamics with Dynamical Lambda}

\author[a,b]{Ruth Gregory}
\author[c]{David Kastor}
\author[c]{Jennie Traschen}

\affiliation[a]{Centre for Particle Theory, Durham University,
South Road, Durham, DH1 3LE, UK}
\affiliation[b]{Perimeter Institute, 31 Caroline Street North, Waterloo, 
ON, N2L 2Y5, Canada}
\affiliation[c]{Amherst Center for Fundamental Interactions, 
Department of Physics, University of Massachusetts, Amherst, MA 01003, USA}

\emailAdd{r.a.w.gregory@durham.ac.uk, kastor@umass.edu, traschen@umass.edu}

\abstract{We study evolution and thermodynamics of a slow-roll transition between 
early and late time de Sitter phases, both in the homogeneous case and in the 
presence of a black hole, in a scalar field model with a generic potential having 
both a maximum and a positive minimum.  Asymptotically future de 
Sitter spacetimes are characterized by ADM charges known as cosmological 
tensions. We show that the late time de Sitter phase has finite cosmological 
tension when the scalar field oscillation around its minimum is underdamped, 
while the cosmological tension in the overdamped case diverges.  We compute 
the variation in the cosmological and black hole horizon areas between the early 
and late time phases, finding that the fractional change in horizon area is 
proportional to the corresponding fractional change in the effective cosmological 
constant. We show that
the extended first law of thermodynamics, including variation in the effective 
cosmological constant, is satisfied between the initial and final states, and 
discuss the dynamical evolution of the black hole temperature.}

\preprint{DCPT-17/23, ACFI-T17-13}

\begin{document}

\maketitle 

\section{Introduction}

The physics of black holes in the early universe is an important subject, about 
which relatively little is known.
The system is both interactive and dynamic, combining effects of cosmic 
expansion with accretion of matter onto the black hole. The gravitational 
thermodynamics is non-equilibrium, involving time dependent areas and 
surface gravities for the black hole and cosmological horizons, such that 
formulating an appropriate definition of temperature proves to be complicated. Still, 
physical implications of primordial black holes have been heavily researched, including
a recent revival of interest that they may provide the dark matter and the progenitors
of the massive black holes detected by LIGO 
\cite{Carr:2009jm, Carr:2016drx, Sherkatghanad:2015rga, Bird:2016dcv, Kawasaki:2016pql, 
Bean:2002kx, Kannike:2017bxn, Domcke:2017fix, Georg:2017mqk,Mediavilla:2017bok}.
In this paper we explore cosmological black holes in a relatively well-controlled 
setting that is also of physical interest, namely the evolution of black holes in slow-roll inflation.
We consider transitions between distinct early and late time  de Sitter phases, 
with evolution driven by a scalar field rolling slowly between a maximum and 
minimum of its potential, both of which are assumed to be positive. The initial 
and final states are described by Schwarzchild-de Sitter (SdS) metrics 
with different values of the black hole mass and cosmological constant.

We begin in \S \ref{sec:nobh} by analyzing pure vacuum to vacuum transitions, 
without a black hole present. Exact results are obtained for an `engineered' example and
compared with the results of a perturbative slow-roll calculation. The growth 
of the cosmological horizon between the two de Sitter vacua is found to obey 
the extended first law \cite{Kastor:2009wy,Dolan:2013ft}, which in the absence 
of a black hole is given by
\be  
T\delta S=\vcal \delta P  
\ee
where the pressure $P$ is provided by the effective cosmological constant,
$\vcal$ is the thermodynamic volume, and $T$ is the initial temperature of the de Sitter horizon. 
Hence $ \delta P $ is dynamically generated by the scalar field. 
Depending on the parameters of the scalar field potential 
there are two qualitatively distinct ways that the metric can approach the late 
time de Sitter metric. The motion of the scalar field
is either overdamped or underdamped as it settles into the true vacuum, corresponding to a
slow-roll or an oscillatory relaxation  respectively. Although in both cases the 
metric decays exponentially fast to de Sitter,
the ADM cosmological tension \cite{Kastor:2016bnm}
is infinite for the overdamped case, but finite for the oscillatory evolution. 
This behavior is analogous to behavior found for AdS black holes with scalar 
fields, as well as in studies of AdS domain-wall/cosmology dualities
\cite{Sudarsky:2002mk, Henneaux:2002wm, Hertog:2004dr, Martinez:2004nb, 
Radu:2004xp, Winstanley:2005fu, Faedo:2015jqa, 
Anabalon:2015xvl, Hertog:2004ns, Henneaux:2006hk, Kaplan:2014dia, 
Skenderis:2006jq, Bazeia:2007vx, McFadden:2010na, Bzowski:2012ih, Kol:2013msa}
in which a scalar field gets a negative mass-squared
from resting at the maximum of a potential, and back-reaction generates an infinite  ADM mass.

In the second part of the paper, we add a black hole to the cosmology and
solve for the evolution of the scalar field and metric in a perturbative ``slow-roll'' 
approximation. This generalizes the calculations of \cite{Chadburn:2013mta} to 
potentials in which both the initial and final states are approximately de Sitter, 
and we focus on thermodynamic aspects of the evolution. 
The black hole grows due to accretion of the scalar field, but the 
cosmological horizon is subject to competing influences. 
While the growing black hole tends to pull the cosmological horizon further in, 
the decaying cosmological constant makes it expand. We find that the expansion dominates.
For a potential interpolating between initial and final values of the cosmological 
constant, $\Lambda_i$ and $\Lambda_f$, we find that the change in both the 
black hole and cosmological horizon areas evolve proportionally to 
$|\delta \Lambda |= |\Lambda_f-\Lambda_i|$ times a factor that depends on 
properties of the Schwarzchild-de Sitter (SdS) spacetime. This is summarized
in equation \eqref{deltaageom} below.
We then show that the extended first law of thermodynamics 
\cite{Kastor:2009wy,Dolan:2013ft}, that
relates the sum of $T\delta S $ contributions from each horizon to 
$\vcal \delta P  $,  is satisfied between the initial and final SdS phases.

The area growth calculations only depend on the temperatures of the 
background static spacetime, but one would like to go further
and work with a dynamical temperature. As a first step,
we utilize a definition of the dynamical black hole temperature $T_{dyn}$ 
based on the Kodama vector \cite{Hayward:2008jq}. This is very nearly
that of a Schwarzschild-de Sitter black hole with the instantaneous 
values of the effective cosmological constant and
black hole radius. We show how the black hole temperature relaxes 
from its initial value to its final asymptotic value, at a rate dependent on the
rolling of the scalar and the local horizon surface gravity. In section
\S \ref{sec:example}, we demonstrate this explicitly for a simple potential, finding
analytic expressions for the dynamical area and temperature.

\section{Pure de Sitter to de Sitter flows}
\label{sec:nobh}

Our goal in this paper is to investigate the effect of a dynamically generated 
cosmological constant on the growth of black hole and 
cosmological horizons, and to explore the thermodynamic relations for 
such evolving black hole systems.
We begin by examining a pseudo-de Sitter spacetime, where the cosmological 
constant varies in time, with no black-hole present. 
As noted in \cite{Skenderis:2006jq}, this is
a double-analytically continued version of an AdS flow, thus an analogue of
the C-theorem \cite{Girardello:1998pd,Freedman:1999gp,Skenderis:1999mm}
tells us that the cosmological constant must always flow to
lower values in time.  
In accordance with this, we consider a real scalar field $\phi$ with
potential $W(\phi)$ and assume that $W(\phi)$ has a maximum $W_i$ 
at $\phi=\phi_i$, and a minimum, $W_f$ at $\phi=\phi_f$.   
If the scalar field starts off at $\phi_i$ at early times, and rolls
to $\phi_f$ at late times, the cosmological constant $\Lambda$ will make a 
transition between the values $W_i/M_p^2$ to $W_f/M_p^2$, 
where $M_p^2 = 1/8\pi G$, at early and late times. 

We take the action for the coupled Einstein plus scalar field field system to be 
\be
S = \frac12\int d^4x\sqrt{-g}\left(-M_p^2R+(\nabla\phi)^2-2W(\phi)\right)
\ee
and use a mostly minus signature.
Assuming an FRW form for the metric $ds^2 = d\tau^2 - a^2(\tau) d{\bf x}^2$,
the equations of motion for the system are given by
\be\label{eom}
\beal
&\left ( \frac{\dot a}{a} \right )^2 = \frac{1}{3M_p^2} \left [
\frac12 {\dot\phi}^2 + W(\phi) \right]\\
&{\ddot\phi} +  3 H {\dot \phi} + \frac{\partial W}{\partial\phi} =0
\eeal
\ee
where $H= \dot{a} /a$. For the pure dS-dS flow, we can simply numerically
integrate these FRW equations for any desired potential, however for the 
analysis of the black hole set-up, it is useful to have analytic solutions, or
approximate solutions, to use to explore the dynamical evolution of the horizons.

\subsection{Engineered flow}
\label{subsec:engineered}

As a first method, we start with an interpolating Ansatz for the scalar field,
and use the Hamilton-Jacobi formulation \cite{Skenderis:2006rr} to engineer 
a potential $W(\phi)$ 
that corresponds to this flow. This approach has been used, for example, 
to generate smooth analytic domain wall solutions in the presence of
gravity \cite{Kehagias:2000au}, and is very similar to the ``fake supersymmetry''
approach described in \cite{Skenderis:2006fb}.

Briefly, the Hamilton-Jacobi approach uses $\phi$ as a time coordinate, 
writing $H = H(\phi)$.  The equations of motion \eqref{eom} then imply that
\be\label{Hprime}
H' = - \frac{\dot{\phi}}{2M_p^2} 
\ee
where the superscript prime denotes a derivative with respect to $\phi$. 
With this identification, the Friedmann equation takes the form of a first order NLDE for $H(\phi)$
\be
H^{\prime2} - \frac{3H^2}{2M_p^2} + \frac{W(\phi)}{2M_p^4}  = 0
\label{HJNLDE}
\ee
A judicious choice of evolution for $\phi(\tau)$ that allows $\dot\phi$ to be
re-expressed as a function of $\phi$ then gives an $H'$ that can be integrated
up to give $H$ and hence $W(\phi)$ using \eqref{HJNLDE}.

For example, if we suppose that the flow of the scalar is
\be\label{phieng}
\phi (\tau)= \eta \tanh (\sqrt{\lambda}\eta\,\tau) 
\ee 
then we find
\be
H' = -\frac{\sqrt{\lambda}}{2M_p^2} (\eta^2-\phi^2) \quad \Rightarrow \qquad
H  (\phi) = H_0 - \frac{\sqrt{\lambda}}{6M_p^2}\phi(3\eta^2- \phi^2)
\label{Heng}
\ee
where $H_0$ is an integration constant. The Hamilton-Jacobi equation 
\eqref{HJNLDE} then finally determines the scalar field potential to be
\be\label{engpot}
W(\phi) = 3 M_p^2 \left ( H_0 - \frac{\sqrt{\lambda}}{6M_p^2} \phi 
(3\eta^2 - \phi^2) \right )^2 - \frac{\lambda}{2} (\phi^2-\eta^2)^2
\ee
Taking $H_0>0$, the potential $W(\phi)$ has a lcoal
maximum at $\phi = -\eta$ and a minimum at $\phi = \eta$.
Correspondingly, the Hubble parameter makes a transition from 
a larger value $ H_i$ associated with the higher vacuum energy
at early times, to a smaller value $H_f$ for the lower vacuum energy at late times, given by
\be\label{hi}
H_i = H_0 + \sqrt{\lambda}\eta^3 /3M_p^2 \  , \  \quad
H_f = H_0 - \sqrt{\lambda}\eta^3 /3M_p^2 ,
\ee
Integrating the expression for $H(\tau)$, obtained by plugging \eqref{phieng} 
into \eqref{Heng}, yields  the scale factor
\be
\log a(\tau) = H_0 \tau - \frac{\eta^2}{3M_p^2} \log\cosh (\sqrt{\lambda}\eta\tau )
+ \frac{\eta^2 }{12M_p^2} \sech^2( \sqrt{\lambda}\eta\tau )
\label{aoftHJ}
\ee
We see that for $|\tau| \gtrsim 1/\sqrt{\lambda}\eta$, $\ln a$ is approximately
linear in $\tau$ with the appropriate Hubble constants given by \eqref{hi}, thus
$\ln a$ is a smoothed out step function, as we would expect.

In the late time limit, the cosmological scale factor and scalar 
field are approximately given by
\be\label{lateaeng}
a(\tau ) \simeq  K e^{H_f \tau } \left (1- \frac{\eta^2 }{2M_p^2} 
e^{- H_f \Gamma\tau } \right) \ \  , \quad
\phi \simeq \eta\left( 1-2 e^{-H_f \Gamma \tau /2 } \right)
\ee
where $K= 2^{(\eta^2/3M_p^2)} $ and we have defined  
$\Gamma H_f =4 \sqrt{\lambda}\eta $ in order to facilitate comparison 
to the late time behavior in subsequent examples. 
Note that with this notation, the expression for the scalar field \eqref{phieng} 
becomes $\phi(\tau)  = \eta \tanh (H_f \Gamma \tau/4) $.   
For fixed $H_f$ the parameter $\Gamma$ then allows one to interpolate between 
slow-roll behavior for $\Gamma\ll 1$ and a sudden change for $\Gamma\gg 1$.

We are  interested in the evolution of the future cosmological horizon. If a light 
signal is emitted at time $\tau$, then as the reception time goes to
infinity the signal is received at a co-moving coordinate separation $r_c (\tau )$ away
\be
r_c  (\tau )=  \int_\tau^\infty \frac{d\tau '}{a(\tau ')}
\label{rcos}
\ee
The cosmological horizon radius $d_c(\tau)$ is the corresponding proper distance
\be\label{dcos}
d_c (\tau ) = a(\tau ) r_c (\tau )\,.
\ee

\begin{figure}
\centering
\includegraphics[scale=0.75]{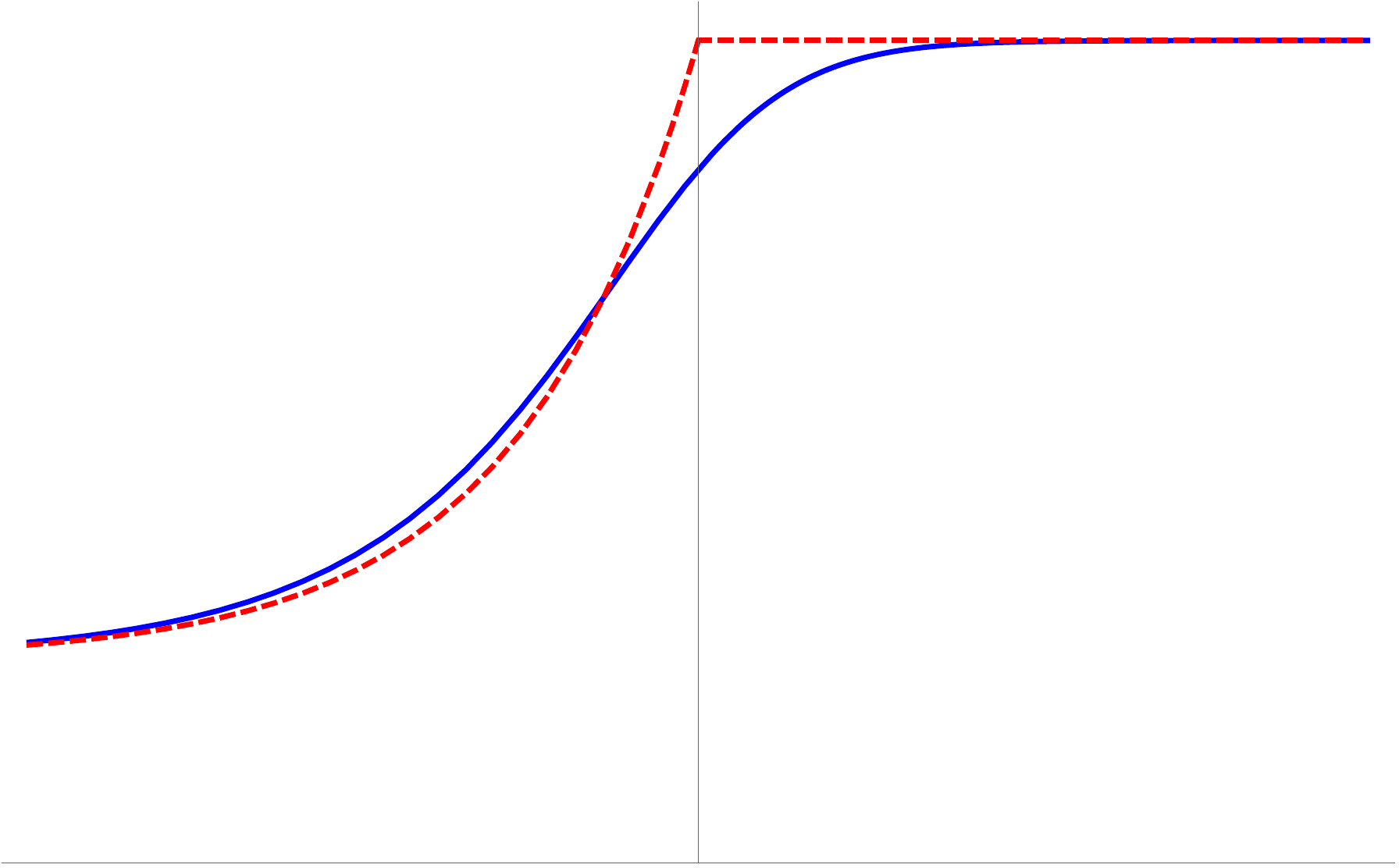}
\caption{The actual cosmological horizon evolution for the Hamilton-Jacobi
engineered potential (blue) vs.\ the approximation (red-dashed).}
\label{fig:hors}
\end{figure}

Evaluating the integral \eqref{rcos} for the cosmological horizon radius $d_c(\tau)$ 
numerically with the scale factor \eqref{aoftHJ} gives a smooth evolution between the 
initial Hubble horizon, $H_i^{-1}$, and the final horizon, $H_f^{-1}$, depicted
in Figure \ref{fig:hors} by a blue line.
It is interesting to compare this exact result to the approximate solution for $r_c$ 
which is gotten by using a step-function approximation for $a(\tau)$,
\be\label{approxa}
a(\tau ) \approx  e^{H_i\tau} \Theta(-\tau) +  e^{H_f\tau} \Theta(\tau)
\ee
Performing the integral  \eqref{rcos} in this case gives an
approximate expression for the evolution of the cosmological horizon radius
\be\label{rcosapprox}
d_c(\tau) \approx   \left[ {1\over H_i } + (\frac{1}{H_f}-\frac{1}{H_i}) 
e^{H_i \tau} \right]\Theta(-\tau)  +  \frac{1}{ H_f }\Theta(\tau)
\ee
Figure \ref{fig:hors} shows the approximate expression compared to the exact one.  
We see in both the exact and approximate results that the cosmological horizon 
interpolates between $d_c = 1/H_i$ at early times and $d_c =1/H_f$ at late times.  
These correspond respectively to the Killing horizons of the early and late time de 
Sitter phases, for the Killing vectors $\xi_* = (\partial /\partial \tau) - H_* \sum_j x^j 
(\partial / \partial x^j)$, where $H_* = H_i$ or $H_f$ respectively.  It is 
satisfying to see that even the simple step-function approximation for $a(\tau )$ 
captures the teleological behavior of the horizon
as $\tau$ increases to the transition time $\tau =0$.

It is interesting to look at the change in the cosmological horizon area 
$A_c=4\pi d_c^2$ over the evolution from early to late times, which is given by
\be\label{changecosa}
\delta A_c= 4\pi({1\over H_f^2}-{1\over H_i^2})
\ee
An extended first law for de Sitter black holes, including variation in the 
cosmological constant was derived in \cite{Kastor:2009wy,Dolan:2013ft}.  
For the case of no black hole (see also \cite{Frolov:2002va}), this reduces to 
\be\label{firstlaw}
T_c\delta S_c=\vthermo\delta P
\ee
where $T_c$ is the temperature of cosmological horizon, $S_c=A_c/4$ is its 
entropy, $P=-\Lambda/8\pi$ is the cosmological pressure, and $\vthermo$ is 
called the thermodynamic volume.  The extended first law, in this simple case, 
relates the change in cosmological horizon entropy to the change in cosmological 
constant.  For a de Sitter spacetime the thermodynamic volume is equal to 
\be\label{cosmov}
\vthermo = {4\pi d_c^3\over 3}
\ee
{\it i.e.}\ the volume of a Euclidean sphere of radius $d_c$, while the horizon 
temperature is $T=1/(2\pi d_c)$.   If we consider a limiting case of our evolution 
such that $H_f=H_i+\delta H$, where $\delta H/H_i\ll 1$, then the change in 
horizon area \eqref{changecosa} is  given to leading order  by 
\be
\delta A_c = -{8\pi \delta H\over H_i^3}
\ee
and it is easily verified that the first law \eqref{firstlaw} is satisfied.
In \S \ref{sec:dynTD} we will see that the first law for a slow-roll inflationary spacetime 
with a black hole also obeys the appropriate extension
of \eqref{firstlaw}.

\subsection{Slow-roll analysis}
\label{subsec:slowroll}

In the previous subsection we assumed a form for $\phi$ that interpolated 
between two de Sitter phases, and found the exact potential and scale factor, 
such that the cosmological evolution was a prescribed flow from the maximum 
of the potential to the minimum. The approach to the true vacuum was overdamped
rather than oscillatory, corresponding to a slow-roll evolution. In this section we 
show that if one starts with a classic double well
potential, with positive minima, then the behavior of the scalar field is 
qualitatively the same as that previously assumed. 

For an analytic treatment we take the standard slow roll assumption 
\cite{Liddle:1994dx} that the scalar field evolution is friction dominated, so that 
\be\label{slowphi}
3H\dot{\phi} \simeq - W'(\phi) \,.
\ee
Now define slow-roll parameters
\be\label{srparam}
\varepsilon (\phi ) =  \frac{M_p^2}{2} \frac{W'^2}{W^2} \ll 1\quad , \qquad
\Gamma (\phi )  = 2 M_p^2 \frac{W''}{W} \ll1
\ee
The second parameter $\Gamma$ (related to the standard slow roll ``eta'' parameter
by a factor of two) is relevant for evolution near a minimum of $W$, where $\varepsilon =0$. 
While $\phi$ is rolling slowly, it will be approximately linear in cosmological time, although
the behaviour near each critical point will be modified. 

Now let $W$ be a double well potential
with a max at $\phi =0$ and a min at $\phi = \eta$,
\be\label{classicv}
W(\phi) = W_f \left( 1+  {\Gamma \over 16\eta^2M_p^2} \left(\phi^2 -\eta ^2 \right)^2  \right)
\ee
where $\Gamma =\Gamma (\eta )$ is the slow-roll parameter evaluated at the 
minimum. The Einstein constraint equation evaluated
at late times, when $\phi = \eta$, gives the standard relation between the 
effective cosmological constant $\Lambda_f = W_f/M_p^2 $ and 
the Hubble parameter in the final de Sitter phase,
\be\label{hv}
H_f^2 =\frac{W_f }{3M_p^2}
\ee

The parameter $\varepsilon \sim \eta^2 \Gamma^2/M_p^2$, 
and we assume that $\Gamma, \varepsilon \ll 1$. Since the 
time dependent part of $\phi$ is already first order, in its equation of 
motion \eqref{slowphi} we can approximate $H\approx H_f$.
The solution for $\phi$ is then 
\be\label{phisr}
\phi^2 = {\eta^2 \over 2} {\rm sech} \left( \frac{H_f \Gamma}{4} \tau\right)
e^{H_f \Gamma \tau/4}
= {\eta^2 \over 2}  \left[ 1 + \tanh \left(\frac{H_f \Gamma}{4} \tau\right)   \right]
\ee
Here an integration constant determining the transition point has been set to 
zero, but one has the freedom to replace $\tau$ in \eqref{phisr}
with $(\tau - \tau_0 )$. 
Hence we see that for the double well potential \eqref{classicv} the scalar 
field has a $\tanh(H_f \Gamma \tau ) $ type behavior, similar to  the engineered case.
This makes sense since the potentials \eqref{engpot} and \eqref{classicv} have 
the same qualitative features for evolving from the maximum to a minimum.
At late times the field goes like
\be\label{latephisr}
\phi \simeq \eta\left( 1-{1\over 2} e^{-H_f \Gamma \tau /2 } \right)
\ee
which has the same decay rate to the true vacuum as in the previous 
example \eqref{lateaeng}. The scale factor
approaches its final de Sitter form with corrections that fall off like $e^{-H_f \Gamma \tau } $.

\subsection{Cosmic hair}

As shown in \cite{Kastor:2016bnm}, asymptotically future de Sitter (AFdS) 
spacetimes carry cosmic hair, encoded in the exponential fall-off terms in the 
metric, analogous to the ADM charges of a black hole.  For black holes, the 
ADM integrals are computed at spatial infinity, while for AFdS spacetimes, 
these integrals are computed at future infinity.  With the AFdS boundary 
conditions established in \cite{Kastor:2016bnm}, the exponentially small 
size of the corrections to de Sitter in this regime are compensated by the 
exponentially growing spatial volume, giving finite results for what were 
referred to as cosmological tension charges\footnote{Cosmic hair in the 
form of cosmological tension charges is fully consistent with the results 
of reference \cite{Wald:1983ky}, even though this work is often mischaracterized 
as a ``cosmic no-hair theorem''.}.  While {\it e.g.}\ the ADM mass and angular 
momentum of a black hole are associated with asymptotic time translation 
and rotation symmetries, cosmological tension is associated with asymptotic 
spatial translation symmetries.  It is analogous to the ADM spatial tension 
charges defined for black brane spacetimes in \cite{Traschen:2001pb} 
(see also \cite{El-Menoufi:2013pza} for the asymptotically AdS case).  
It was found \cite{Kastor:2016bnm} that cosmological tension in a given 
spatial direction captures the leading order correction of the scale factor in 
that direction to its limiting late time de Sitter behavior.

In this paper, we consider transitions between early and late time de Sitter 
phases, and can compute the cosmological tensions associated with the 
approach to the late time de Sitter limit.  Assuming that the minimum of the 
potential is at a finite value of $\phi$, we find that the cosmological tensions 
will be finite, if the approach to the minimum is underdamped.  However, in the overdamped case, the corrections to 
de Sitter behavior fall off too slowly, and the cosmological tension charges diverge.

In both the engineered example in \S \ref{subsec:engineered} and the approximate 
analytic solution for the slow-roll example in \S \ref{subsec:slowroll},  the scalar 
field and metric indeed decay to the late time de Sitter vacuum 
exponentially fast. This agrees with expectations based on \cite{Wald:1983ky},
which showed that an initially expanding, homogeneous spacetime with 
cosmological constant $\Lambda>0$, matter fields satisfying the 
weak ($\rho \geq 0$) and strong ($\rho + 3p \geq 0$) energy conditions, 
and non-positive spatial curvature, falls off to de Sitter exponentially quickly at late times. 
To align our scalar field system with the assumptions of \cite{Wald:1983ky}, 
we take the cosmological constant to be given in terms of  the final value of the 
scalar potential $\Lambda = W_f/M_p^2$.  The scalar field then moves in the shifted potential 
\be
U(\phi) = W(\phi)-W_f \ge 0
\ee
For the scalar field evolution the energy density and pressure are given by 
\be
\rho = {1\over 2}\dot\phi^2 +U(\phi)\ge 0,\qquad p = {1\over 2}\dot\phi^2 -U(\phi)\ge 0
\ee
Since $\rho\ge 0$ the weak energy condition is satisfied. However, 
$ \rho +3p = 2(\dphi ^2  - U(\phi) )$, which is negative in the early time de 
Sitter phase, so that the strong energy condition is not satisfied.  
Nonetheless, we find that the metric
and scalar field still approach the late time de Sitter vacuum exponentially fast.
This decay is generic, determined by the fact that near a minimum of the 
potential the wave equation for $\phi$ becomes that of a damped 
simple harmonic oscillator, with damping provided by the Hubble expansion. 
However, we show that in the overdamped case the 
detailed decay rate is too slow to result in a finite cosmological tension charge.

Assume that the scalar field potential $W(\phi)$ has a maximum and a 
minimum so that there can be two de Sitter phases as the field
evolves from an initial value $\phi_i$ at the maximum $W_i$ to a final 
value $\phi_f$ at the minimum $W_f$, as in the two examples above.
We want to solve equations  \eqref{eom} for $\phi(\tau)$ and $a(\tau )$ 
at late times as  $\phi$ approaches the minimum at $\phi_f$.  
Assume that the second derivative of the potential is non-zero at $\phif$, 
so that near the minimum the scalar field potential can be approximated by
\be\label{expandv}
W(\phi ) \simeq W_f + {1\over 2}\ddvf  \left(\phi - \phif \right)^2 
\ee
To leading order the wave equation \eqref{eom} for $\phi$ then reduces to the 
ODE for a damped simple harmonic oscillator
\be\label{sho}
\ddot{\phi } + 3H_f \dphi +\ddvf  \left(\phi - \phif \right) = 0
\ee
which has the solutions 
\be\label{phigen}
\phi(\tau)  = \phi_f+ \phi_1 e^{-\beta_\pm H_f  \tau} \  , \quad 
\beta_\pm = {3\over 2} \left( 1\pm \sqrt{ 1-2\Gamma /3 } \right) 
\ee
where $\phi_1$ is a constant and $\Gamma  =2\ddvf / 3H_f^2 $. Overdamped 
oscillation results from real values of $\beta$,  corresponding to  
$\Gamma \leq 3/2$, and in this case $\beta_- $ is the dominant mode. 
From \eqref{srparam} we see that slow-roll evolution corresponds to $\Gamma \ll 1$, giving
\be
\phi = \phi_f+ \phi_1 e^{-\Gamma H_f  \tau /2} \   \quad \text{(slow-roll)}
\ee
in agreement with the late time slow-roll evolution in \eqref{latephisr}.  

The late time behavior of $\phi$ determines the late time behavior of the scale factor.
As $\phi$ approaches its value at the minimum of the potential $W(\phi)$  
the Einstein constraint equation becomes
\be\label{constraint} 
\left({\da \over a} \right)^2  \simeq
\frac{1}{3M_p^2}\left (W_f +  {1\over 2}\ddvf  \left(\phi - \phif \right)^2 + {1\over 2} \dphi ^2\right )
\ee
the evolution of the scale factor in this regime can be written as 
\be\label{latea}
a(\tau ) \simeq e^{H_f \tau } +\delta a(\tau)
\ee
where the deviation $\delta a(\tau)$ from the late time de Sitter evolution is small.
The leading terms relate the late time value of the Hubble parameter to the value 
of the potential at its minimum, as in \eqref{hv}, while the next order terms give 
\be\label{deltaa}
\delta a(\tau) =-{\phi_1^2 \over  24M_p^2 \beta_- }  \left( \beta_-^2 H_f^2 
+ \ddvf  \right) e^{H_f (1- 2\beta_- )\tau }
\ee
In the slow-roll approximation, with $\Gamma \ll1$, the late time limit of the scale factor is then
\be\label{deltaaslow}
a (\tau)= e^{H_f \tau } \left( 1   - \frac{\phi_1^2H_f^2}{8M_p^2}  e^{- \Gamma H_f   \tau }  \right)  \   
\quad \text{(slow-roll)}
\ee
in agreement with \eqref{lateaeng}. 
Critically damped motion occurs when $\Gamma =3/2$. In this case the two roots 
coincide, and the second linearly independent mode
goes like $ e^{-\beta H_f \tau} \ln \tau $. 

Underdamped oscillations occur when $\Gamma > 3/2$ \cite{Skenderis:2006rr}, 
and at this point there is a 
qualitative change in the aymptotic behavior of $\phi$ and $a(\tau)$
as the fields oscillate around the minimum with decaying amplitude,
\be\label{deltaafast}
\beal
\phi(\tau)  &  = \phi_f+\phi_1 e^{-3 H_f  \tau /2} \sin \omega \tau \\
a & = e^{H_f \tau } \left( 1  -  e^{- 3  H_f   \tau } \left[ \frac{1}{36M_p^2} \phi_1^2 \ddvf      
+ C_1 \cos 2 \omega\tau + C_2 \sin 2 \omega\tau \right] \right) 
\eeal
\ee
where $ \omega^2 =\ddvf - H_f^2 $, and $C_1$ and $C_2$ are constants whose precise 
expressions will not be needed.
The underdamped, oscillatory, cases all decay at the
same rate, which is precisely that needed for a finite, non-zero, cosmological  
tension\footnote{Since the metric is isotropic, all three tensions, associated with 
invariance in the three spatial directions, are the same.} 
$\cal{T}$.
The ADM charges are defined in terms of  boundary integrals, which require a number of further definitions in order to specify.  We present this material in a short appendix to the paper.
In computing $\cal{T}$ we average over a period, and the oscillatory terms average to zero. 
For the non-oscillatory solutions \eqref{deltaa}, the key ingredient  in the boundary integral is the behavior of a term like 
\be\label{btterm}
{\delta \da \over a_{ds} }dv \simeq e^{H_f (3-2\beta_- ) \tau}
\ee
where $dv  = e^{3H_f \tau}$ is the area element on a constant
time slice of the background de Sitter metric. As $\tau \rightarrow \infty$ this is finite 
and non-zero only for the critically damped case of $\beta_- =3/2$, whereas
the integral diverges for all the overdamped cases which have $\beta_-<3/2$.
However, in the critically damped case the two modes corresponding to 
$\beta_\pm$ coincide, and the dominant late time 
mode is instead a second linearly independent solution that goes like 
$(\ln \tau) e^{-3H_f \tau /2}$ and
also leads to a divergent  tension. Hence it is only the oscillatory modes 
that yield a finite cosmological tension ${\cal T}$.
Since the tension involves an integration over an infinite spatial area, 
we make the spatial coordinates
periodic with period $L$, and the finite quantity is the tension per unit area. For the underdamped case, where the tension is finite, one then finds
\be\label{tension}
\frac{2{\cal T}}{M_p^2 L^2 } ={9 \over 4}  \phi_1^2  H_f 
\ee
where $\phi_1^2$ sets the magnitude of $\phi -\phi_f $ and $\delta a$.

To summarize, there are two qualitatively distinct ways for $\phi$ to decay to  
the minimum of the potential, either with or without oscillations. 
These behaviors are distinguished by finite or infinite cosmological tension. 
Note that if $\phi$ were coupled to other fields to reheat the universe, this would 
also distinguish different physical mechanisms
for the reheating. 

Lastly, It is interesting to see how these different asymptotics look in AdS.
In four dimensions the metric of a planar AdS black hole approaches AdS at spatial 
infinity exponentially fast,  like $e^{-3y/\ell}$,
where $y$ is a proper length radial coordinate and $\ell$ is the AdS length scale. 
However, the relevant area element grows like $e^{3y/\ell}$, 
so the resulting boundary term for the ADM mass is finite. A significant amount of 
research has been done on the behavior of  scalar fields in AdS/ CFT, and
particularly analogous to the issue of future asymptotically  de Sitter  
spacetime are studies of  AdS black holes with scalar hair, AdS domain-wall
scenarios, and holographic domain walls \cite{Sudarsky:2002mk, Henneaux:2002wm, 
Hertog:2004dr, Martinez:2004nb, Radu:2004xp, Winstanley:2005fu, Faedo:2015jqa, 
Anabalon:2015xvl, Hertog:2004ns, Henneaux:2006hk, Kaplan:2014dia, 
Skenderis:2006jq, Bazeia:2007vx, McFadden:2010na, Bzowski:2012ih, Kol:2013msa}.
In these situations the potential for the scalar field is negative, and the field 
approaches a maximum of the potential, which provides
an effective negative mass-squared. 
The two modes of the field decay like $e^{-\Delta_\pm y/\ell}$,
\be\label{adsdecay} 
\Delta_\pm = {3\over 2} \left( 1\pm \sqrt{ 1+4 m^2 \ell^2 /9 } \right)  \ , \quad 
m^2 = W^{\prime\prime}_\infty < 0
\ee
with the constraint $m^2_{BF} \leq m^2 \leq m^2_{BF} +1 <0$, where 
$m^2_{BF} = -9/(4 \ell^2 )$ is the Breitenlohner-Freedman bound 
\cite{Breitenlohner:1982jf,Sudarsky:2002mk, Hertog:2004dr, Martinez:2004nb, Radu:2004xp, 
Winstanley:2005fu, Faedo:2015jqa, Anabalon:2015xvl}.
This formula is analogous to the scalar field decay in the cosmological case 
\eqref{phigen}, with the substitution of a negative potential in which $\phi$
approaches a maximum,  for the positive potential with $\phi$ approaching 
a minimum. The dominant far field mode with $\Delta_-$ decays
too slowly for the spacetime to have a finite ADM mass, as we have found for 
the late time de Sitter behavior with real $\beta_-$. In the AdS case 
a finite ADM charge can be constructed by combining  contributions from 
the gravitational and scalar fields \cite{Henneaux:2002wm, Hertog:2004dr, 
Hertog:2004ns, Henneaux:2006hk}.
A similar situation arises in AdS domain wall spacetimes in which a scalar field 
potential  interpolates in the radial direction between two AdS vacua.
The interpolation between AdS vacua corresponds to an RG flow between two CFTs, and 
has previously been compared to de Sitter to de Sitter transitions \cite{Kaplan:2014dia}.
A topic for future analysis
is to work out the combination of  cosmological tension with a scalar field 
contribution to form a finite generalized ADM tension, appropriate
for a slow roll approach to de Sitter.

\section{Schwarzschild-quasi de Sitter spacetimes}
\label{sec:sectionsds}

In \S \ref{sec:nobh} we analyzed  de Sitter to de Sitter evolutions, with no 
black hole present. We now investigate the effects of a black hole on this 
transition, and in particular how the black hole and cosmological horizons evolve.
Exact solutions of dynamical black holes include the McVittie metric 
\cite{McVittie:1933zz}, which is simply SdS in cosmological coordinates in the 
most physical case, and examples of maximally charged multi-cosmological black 
holes both without  \cite{Kastor:1992nn} and with scalar fields \cite{Gibbons:2009dr}.
Accretion of fields and growth of cosmological black holes has been studied in 
different approximations and numerically, addressing scalar field cosmologies, 
generalizing the properties of Killing black holes, and the dynamics 
of accretion \cite{Jacobson:1999vr, Saida:2000at, Frolov:2002va, Harada:2004pf,
Sultana:2005tp, Faraoni:2007es, Carrera:2008pi, Hayward:2008jq,  Rodrigues:2009eg, 
Carr:2010wk, UrenaLopez:2011fd, Guariento:2012ri,
Rodrigues:2012xm, Abdalla:2013ara, Babichev:2014lda,
Afshordi:2014qaa,Davis:2016avf,Frolov:2017asg}.

In this paper we follow the approach of \cite{Chadburn:2013mta} and systematically apply 
perturbation theory and the slow roll approximation to the Einstein-scalar field equations, 
identify the horizons, and use well-behaved coordinates on the horizons to compute the 
behavior of the scalar field and thermodynamic properties. Our results yield analytic 
expressions for the evolution of horizon areas, temperatures, thermodynamic volume, 
and local pressure. We find that the generalised first law of thermodynamics is satisfied 
between the final and initial SdS states.
Though the results are given in terms of a general potential $W(\phi )$, 
it is most straightforward to make a thermodynamic interpretation in
the case that $W$ has a maximum and a minimum, since then the initial and final
states are equilibria described by a static SdS metric. The main results of this 
section  are summarized in equations \eqref{deltabchsol}, 
\eqref{changecha}, and \eqref{deltabsol}, \eqref{changebha},
which give the total change in the cosmological and black hole horizon areas 
respectively. The reader uninterested in the derivation
can skip to those results without loss of continuity. 
 
For a slowly rolling scalar field, the spacetime will be adiabatically de Sitter (or SdS), 
with small, ``time-dependent'' corrections. The idea therefore is to first find a solution
with a constant $\phi$, then to correct this perturbatively for a rolling $\phi$. 
Given that we have a cosmological evolution in `time', together with the black hole 
giving us a `radial' dependence of our geometry, our analysis should capture the 
dependence on these two coordinates. Since the black hole and cosmological event 
horizons represent coordinate singularities in the standard SdS metric, we follow the 
methodology of reference \cite{Chadburn:2013mta} that treated 
black hole evolution for a scalar with an exponential potential, using 
null coordinates for the metric that encode the dependence on two parameters: 
\be
ds^2 = 4 e^{2\nu}\sqrt{\frac{B_0}{B} } dUdV - Bd\Omega_{{\rm II}}^2 ,
\label{weylform}
\ee
but are nonsingular at the horizons, and indeed facilitate the analysis of horizon
behaviour that are now located at ``$U=0$'' or ``$V=0$''. Here, $B_0^{1/2}$ is
a fiducial length scale that maintains dimensional consistency.
This form of the metric also clearly identifies the main physical degree of freedom of 
the gravitational field as the $B-$function, since $\nu$ communicates with the remaining
gauge freedom of conformal transformations in the $U,V$ plane. In the absence of 
any scalar evolution, the physical degree of freedom in $B$ corresponds to the
mass of the black hole as we now briefly review. (This discussion follows 
\cite{Chadburn:2013mta, BCG}).

In this null gauge, the Einstein equations become
\bea
\phi_{,UV} &=& - W_{,\phi}(\phi)\sqrt{\frac{B_0}{B} }e^{2\nu}
- \frac{1}{2B}\left(B_{,U}\phi_{,V}+B_{,V}\phi_{,U}\right)
\label{phiUVeq}\\
B_{,UV} &=& 2\left(\frac{W(\phi)}{M_p^2} B^{1/2} - B^{-1/2}\right)e^{2\nu}B_0^{1/2}
\label{BUVeq}\\
\nu_{,UV} &=& \frac{1}{2}\left(\frac{W(\phi)}{M_p^2} B^{-1/2}
+ B^{-3/2}\right)e^{2\nu}B_0^{1/2} - \frac{\phi_{,U}\phi_{,V}}{2M_p^2}
\label{nuUNeq}\\
B_{,VV} &=& 2\nu_{,V} B_{,V} - B \phi_{,V}^2 /M_p^2
\label{Vint}\\
B_{,UU} &=& 2\nu_{,U} B_{,U} - B \phi_{,U}^2 /M_p^2
\label{Uint}
\eea
If $\phi$ is constant, equations \eqref{Vint} and
\eqref{Uint} give
\be
2\nu = \log \frac{B_{,V}}{\sqrt{B_0}} + G'(U) = \log \frac{B_{,U}}{\sqrt{B_0}} + F'(V)
\ee
where $F$ and $G$ are in principle arbitrary functions, expressed here
as derivatives for convenience. From this, we deduce that
\be
B = B\left[F(V)+G(U)\right]
\quad \text{and} \quad e^{2\nu} B_0^{1/2} = F'G'B'
\ee
Then \eqref{BUVeq} can be integrated up to give
\be\label{mudef}
B' = \frac{4W_0}{3M_p^2} B^{3/2} - 4 B^{1/2} + \mu
\ee
where the integration constant $\mu$ is nonzero if a black hole is present.
Substituting these expressions back into the metric gives
\be
ds^2 = -16 \left ( 1- \frac{W_0}{3M_p^2} B - \frac{\mu}{4\sqrt{B}} \right )
dFdG - Bd\Omega_{{\rm II}}^2 .
\ee

Comparison with the SdS metric suggests that we identify $\mu = 8GM$,
and choose a radial coordinate $r = \sqrt{B}$. It is then fairly clear that if we
identify $F$ and $G$ with the advanced and retarded null coordinates
\be
F \leftrightarrow -\frac{(t + r^\star)}4\qquad , \qquad 
G \leftrightarrow \frac{(t- r^\star)}4
\label{FGtrstar}
\ee
(where $r^\star$ is the standard tortoise coordinate), then transform to
$t,r$ coordinates we recover the SdS metric: 
\be\label{sdsmetric}
ds^2 = N(r) dt^2 - \frac{dr^2}{N(r)} - r^2 d\Omega_{II}^2\,.
\ee
We have written the SdS potential as
\be\label{sdsn}
\beal
N(r)   = 1 - \frac{2GM}{r} - H_0^2 r^2 
= - \frac{H_0^2}{r} (r-r_c)(r-r_h)(r-r_N),
\eeal
\ee
where $H_0^2 = W_0/3M_p^2$, and we identify the black hole horizon 
(if present) as $r_h$, the cosmological horizon as $r_c$, and the remaining 
zero of $N$ as $r_N = -(r_c+r_h)$. Note, the roots of $N$ are related to
the physical parameters via 
\be\label{sdsparam}
\beal
\Lambda &= 3 H^2 = 3 / (r_c^2+r_h^2+r_hr_c)\\
2GM &= H^2 r_c r_h (r_c+r_h)
\eeal
\ee
and the tortoise coordinate $r^\star$ is given explicitly by
\be\label{rstarbh}
r^\star (r)  = \int \frac{dr}{N(r)} = {1\over 2\kappa_c } \log \left |\frac{r-r_c}{r_c} \right | 
+ {1\over 2\kappa_h }  \log \left | \frac{r-r_h}{r_h} \right | 
+ {1\over 2\kappa_N }  \log \left | \frac{r-r_N}{r_N} \right |\,.
\ee
Here, $\kappa_i$ are the usual surface gravities at the individual horizons,
$2\kappa_i = N'(r_i)$. 

Finally, although $F$ and $G$ are null coordinates, the metric still has 
coordinate singularities at the black hole and cosmological event horizons.
These can of course be removed locally by using the standard
Kruskal coordinates:
\be
\beal
&\bullet \; r\to r_c &\qquad 
V &= {1\over 2\kappa_c } \exp \left [ \kappa_c (t+r^\star)\right] \;,
&\quad
U &= {1\over 2\kappa_c }\exp \left [ \kappa_c (t-r^\star)\right] \\
&\bullet \; r\to r_h &\qquad 
v &= {1\over 2\kappa_h }  \exp \left [ \kappa_h (t+r^\star)\right] \;,
&\quad
u &= - {1\over 2\kappa_h }\exp \left [ -\kappa_h (t-r^\star)\right] 
\eeal
\label{kruskcos}
\ee
however, no global maximal extension of the SdS coordinates is
possible. Given that we are interested in the future evolution of the
spacetime, we could choose to use $\{u,V\}$ as nonsingular 
coordinates, however, in practice it is easier to analyse physics near 
the event horizons in the local Kruskals.

\subsection{The scalar field behaviour}

Prior to starting the analysis of an evolving black hole, it is useful to relate
this notation to our previous discussion with cosmological time. 
With the SdS black hole, the natural solution is expressed in static gauge, 
as we have discussed above, but it is useful to see how the null and static 
coordinates relate to the cosmological coordinates. 

Setting $\mu=0$, and integrating \eqref{mudef} gives
\be
B[X] = \frac{1}{H_0^2} \tanh^2 [-2H_0X]
\ee 
where $X=F+G<0$. Using \eqref{FGtrstar}, \eqref{rstarbh}, 
and $\kappa_c=-H_0= - r_c^{-1}$,
\be
X = -\frac{r^\star}{2} = \frac1{4H_0} \ln \left | \frac{1 - H_0r}{1+H_0r} \right| 
\ee

Taking $V\pm U$ and using \eqref{kruskcos} then relates these 
expressions to our canonical cosmological
coordinates (via conformal time $\hat{\eta} = \int dt/a$) as
\be\label{cosst}
\beal
\tau &= -\frac1{H_0} \ln \left [-H_0(U+V)\right] = t + \frac{1}{2H_0} \log(1-H_0^2 r^2) \\
\rho &= \left ( V-U \right) = \frac{r e^{-H_0 t}}{\sqrt{1 - H_0^2 r^2}}
\eeal
\ee
In the cosmological slow-roll approximation with no black hole, $\phi$ only 
depends on cosmological time $\tau$. Even though this is a more involved
expression in the static gauge, we still have the notion that $\phi$ depends
on a single function of $t$ and $r$. In \cite{Chadburn:2013mta}, it was found
that in the slow-roll approximation for the case of an exponential scalar 
potential, in the presence of a black hole $\phi$ depended linearly on a 
variable $x(t,r)$, which was a generalization of the cosmological time coordinate 
in \eqref{cosst}. 
Here we maintain this notion of slow-roll, and look for a similar $x(t,r)$ that
reduces to $\tau$ as the black hole area goes to zero.

In this slow-roll approximation, derivatives of $\phi$ and $W'(\phi)$ are assumed
to be small quantities relative to the overall magnitude of the potential, thus 
(\ref{BUVeq}-\ref{Uint}) reduce to the pure cosmological constant equations
we have just discussed, and \eqref{phiUVeq} to leading order requires only
these background forms of the metric functions to find the evolution of $\phi$
due to its potential in the presence of the black hole.
Substituting in these forms, \eqref{phiUVeq} is
\be
\frac{\phi_{,FG}}{N(r)} - \frac{2}{r} \left ( \phi_{,F} + \phi_{,G} \right) 
= 4 \frac{\partial W}{\partial\phi}
\label{phiFGeq}
\ee
(recalling that $r$ is a function of $F+G$). 

We now look for a variable $x(t,r)$, such that $\phi$ is predominantly a
function of $x$. That is, $x$ must be  suitably chosen to render \eqref{phiFGeq}
an ODE for $\phi$ when the $\phi''$ term is neglected. Now
\be
\frac{\phi_{,FG}}{N(r)} - \frac{2}{r} \left ( \phi_{,F} + \phi_{,G} \right) 
= \frac{x_{,F} x_{,G} }{N} \phi'' + \left [ \frac{x_{,FG}}{N} - 2\frac{x_{,F}+x_{,G} }{r} \right] \phi' 
\label{phiFG}
\ee
so let
\be
x = t + \xi[r]
\ee
then, 
\be
\beal
x_{,F} &= 2 + \xi' \frac{B'}{2\sqrt{B}} = 2(1-N\xi')\\
x_{,G} &= -2 + \xi' \frac{B'}{2\sqrt{B}} = -2(1+N\xi')
\eeal
\ee
and \eqref{phiFGeq} becomes
\be
\frac{N^2\xi^{\prime2}-1}{N} \phi'' + \frac{(r^2 N \xi')'}{r^2} \phi' 
= \frac{\partial W}{\partial\phi}
\ee

We can now read off our requirement for slow-roll as
\be
\xi' \propto \frac{C+r^3}{r^2N}
\ee
The final determination of the constant of proportionality and the integration
constant $C$ is determined by the boundary conditions that the field is purely ingoing
at the black hole horizon and purely outgoing at the cosmological horizon, leading to 
\be\label{xcoord}
x = t - r^\star +  {1\over \kappa_h} \ln \left | \frac{r-r_h}{r_h} \right | 
+ \frac{r_c }{2\kappa_h r_h} \ln \left | \frac{r-r_N}{r_N} \right |
-\frac{r_h r_c}{r_c-r_h} \ln \frac{r}{r_0} 
\ee
where $r^\star$ is given in \eqref{rstarbh}. The new $x$ coordinate
reduces to the expression for the cosmological coordinate $\tau$ in \eqref{cosst} when $r_h =0$.

This implies that
\be
\xi' = \frac{1}{N} \left ( \frac{r_c^2 r_h^2(r_c+r_h)
- r^3 (r_c^2+r_h^2)}{r^2(r_c^3-r_h^3)} \right)
\label{xiprime}
\ee
and dropping the $\phi''$ term, the slow-roll equation becomes
\be\label{slowroll}
3 \gamma \phi '  (x) = -  \frac{\partial W}{\partial\phi}
\ee
where 
\be\label{gamma}
\gamma =  \frac{ r_c^2+r_h^2 }{r_c^3-r_h^3 } ={A_{tot} \over 3 \vcal} 
\ee
Here $A_{tot}= 4\pi ( r_c^2+r_h^2)$ is the total horizon area, and 
$ \vcal  =4\pi ( r_c^3-r_h^3 ) /3 $ the thermodynamic volume,
of the de Sitter black hole system. The volume $\vcal$, with $r_h =0$, 
arose in the thermodynamics of the cosmological horizon in the absence of a 
black hole, equations \eqref{firstlaw} and \eqref{cosmov}. It will again enter 
the thermodynamics of the black hole$-$cosmological system in subsequent sections.

It is interesting to compare \eqref{slowroll} to the cosmological slow-roll 
equation, where $\gamma = H_0$. Clearly as $r_h\to0$, the two slow-roll equations
coincide, but we can now explicitly see the effect of the black hole 
on the friction for $\phi-$motion. As $r_h$ is switched on, the denominator in 
\eqref{gamma} decreases and the numerator increases, hence $\gamma$
is larger for larger black holes at fixed $\Lambda$.\footnote{It is a little
more subtle, since for fixed $H^2$, {\it i.e.}\ $W$, $r_c$ decreases as $r_h$
increases, however, it is easy to check that the overall effect of increasing
the black hole size is to increase $\gamma$.} Thus the effect of a black
hole is to further slow down the slow roll inflation, and for black holes
very close to the Nariai limit, the evolution becomes arbitrarily slow.
Indeed, expanding $\gamma$ for small and large black holes,
demonstrates this effect clearly:
\be\label{gammatolambda}
\gamma \sim
\begin{cases}
H_0 \left ( 1 + GM H_0 \right ) & M\to0 \\
\frac{2}{3(r_c-r_h)} & r_h \to r_c
\end{cases}
\ee

Finally, given that the coefficient of the $\phi''$ term contains a $1/N$ factor, 
we must check that this term remains small. From \eqref{xiprime}, we see 
that $N\xi' \to \pm 1$, at the black hole and cosmological horizons, thus the 
term multiplying $\phi''$ is in fact regular at the horizons, and thus overall stays small. 
It is therefore consistent to drop the second derivative term as long as
$|\phi'' | \ll \gamma \phi'$.

\subsection{Growth of the event horizons}

What does the resulting evolution look like? Two fundamental geometrical 
properties of the spacetime are the areas of the
black hole and cosmological horizons. Since the black hole is accreting the scalar field,  
we expect the black hole to grow.
A bigger black hole will tend to pull in the cosmological horizon. 
However, the effective cosmological constant is decreasing,
which leads the cosmological horizon to grow. We will see that the second effect 
dominates, and that both horizons grow.

Ideally, we would calculate the full gravitational back reaction throughout the
spacetime, as described in \cite{Chadburn:2013mta}, however, this process
is rather involved, and somewhat specious to the main theme of our discussion,
namely the evolution of the event horizons. As it turns out, this is fairly straightforward to extract.
Note that in the background solution, the horizon (a null surface) is at fixed $r$, 
{\it i.e.}\ fixed $B$. Taking the local Kruskals at each horizon, given in equation \eqref{kruskcos},
the cosmological event horizon is defined as $V=0$, and parametrised by $U$
whereas the black hole event horizon is defined as $u=0$, and parametrised by $v$.
In the vicinity of the horizons of the background SdS we have:
\be
\begin{cases}
V = U\exp \left [ 2\kappa_c r^\star\right] \approx U (r-r_c) 
& {\rm as}\; r\to r_c\\
u = -\frac{1}{4\kappa_h^2 v} \exp \left[ 2\kappa_h r^\star \right] 
\approx -\frac{1}{ 4\kappa_h^2 v}  (r-r_h) & {\rm as} \; r\to r_h
\end{cases}
\ee

To get the horizon growth, the idea is to expand the Einstein equations \eqref{Vint} for the 
black hole horizon, and \eqref{Uint} for the cosmological horizon, 
by using the fact that $B_{,u}$ ($B_{,V}$)  is zero on the 
black hole (cosmological) event horizon in the background spacetime.
\bea
\bullet \; r\to r_c \qquad\qquad
\delta B_{,UU} &=& 2\nu_{,U} \delta B_{,U} -  B \phi_{,U}^2 /M_p^2
\label{cospert}\\
\bullet \; r\to r_h \;\qquad \qquad
\delta B_{,vv} &=& 2\nu_{,v} \delta B_{,v} - B \phi_{,v}^2 /M_p^2
\label{bhpert}
\eea
Here the derivative of $\phi$ terms enter as a perturbative source.
Lastly, we will need the expression for $\nu$ at each horizon:
\be\label{nearhorcoords}
\beal
&\bullet \; r\to r_c &\qquad 
e^{2\nu} &=  {r N(r) \over 4\kappa_c^2 UV}
\approx -  \frac{r_cN}{ 4\kappa_c^2 U^2 (r-r_c)}
\approx - \frac{r_c}{ 2\kappa_c U^2}\\
&\bullet \; r\to r_h &\qquad 
e^{2\nu} &= - {r N(r) \over 4\kappa_h^2 uv }
\approx  \frac{r_hN}{(r-r_h)} \approx  2\kappa_h r_h
\eeal
\ee

\subsubsection{Cosmological event horizon}

Starting with the cosmological event horizon, as $r\to r_c$, 
$x\sim  \kappa_c^{-1} \ln (2\kappa_c U) + $const., and from \eqref{nearhorcoords} 
we have $\nu_{,U} \simeq -1/U$ along the 
cosmological horizon. Hence \eqref{cospert} gives
\bea
\left ( U \delta B \right)_{,UU} 
&=& -\frac{r_c^2}{\kappa_c^2 M_p^2}  \frac{ \phi^{\prime 2}}{U}
= -\frac{ r_c^2}{ \kappa_c M_p^2} \phi^{\prime 2} \frac{d x}{dU}\\
\Rightarrow \quad \left ( U \delta B \right)_{,U} 
&=& -\frac{ r_c^2}{ \kappa_c M_p^2} \int \phi^{\prime 2} {d x} 
= \frac{ r_c^2}{3\gamma \kappa_c M_p^2} \int {\partial W \over \partial \phi} d\phi \nonumber\\
&=& -\frac{ r_c^2}{3\gamma \kappa_c M_p^2}
\left (  W_i - W\left[ \phi(U)\right]  \right )
\eea
Here we have set  $\delta B =0$ when $\phi$ takes its initial value $\phi_i$, 
which is as $U\rightarrow -\infty$ and $W_i =W(\phi_i) $.
Integrating again gives
\be\label{deltabchsol}
\delta B = -\frac{ r_c^2  }{3\gamma | \kappa_c | M_p^2 U} \int_U^0
\left (W_i -  W[\phi(U^\prime )] \right) dU^\prime
\ee
where we have written $-\kappa_c = |\kappa_c |$ to clarify that the change in the 
horizon area is positive. As $U\to-\infty$, $\phi\to \phi_i $, and we get $\delta B \to 0$, 
but as we go to future infinity, or $U\to0$ ,  then $\phi \to \phi_f $ and
\be
\delta B = \frac{1}{3 \gamma |\kappa_c  |M_p^2} r_c^2  (W_i - W_f ) \  ,
\ee
This means that the total change in cosmological horizon
radius is positive, and given by
\be
\delta r_c =\delta \sqrt{B(r_c)} = \frac{\delta B}{2r_c} 
=\frac{1}{6 \gamma |\kappa_c | M_p^2 } r_c  (W_i -W_f )
\label{deltarc}
\ee
When there is no black hole, then $r_h=0$, $r_c  = 1/H$, $| \kappa_c |=  H$, 
and the change in horizon radius is
$H_f^{-1}-H_i^{-1}$, in agreement with equation \eqref{rcosapprox} and Figure \ref{fig:hors}.
 
Finally, this gives for the change in the cosmological horizon area 
$\delta A_c = 8\pi r_c \delta r_c $ 
\be\label{changecha}
\delta A_c = \frac{A_c}{ 3\gamma |\kappa_c | M_p^2} (W _i - W_f )
\ee
This can be written in terms of the change in the early and late time effective 
cosmological constants by using $\Lambda_I =  W_I/M_p^2$ .
Since $W_i > W_f$,  corresponding to evolution of $\phi$ by rolling down the 
potential, the cosmological horizon grows. 
 
It is of interest to find how much the presence of a black hole affects the 
growth of the cosmological horizon, for a fixed potential. In equation \eqref{changecha} 
the quantities $A_c ,\  \kappa_c $ and $\gamma$ all depend on $r_h$. 
Looking at the limits, the effect is of course negligible for
very small black holes, but as the size of the black hole horizon becomes 
comparable to that of the cosmological horizon, 
the total change in $A_c$ is diminished by a factor of approximately two-thirds. 

\subsubsection{Black hole event horizon}

The analysis for the black hole horizon area is similar to that for cosmological horizon, 
but now we use the Kruskal coordinates $u,v$. The black hole horizon is at $u=0$, and 
$v$ is the coordinate along the horizon. As $r\rightarrow r_h$, we have 
$x\sim \kappa_h^{-1} \ln (2\kappa_h v) + $const. From equation \eqref{nearhorcoords} it follows that
$\nu$ is constant on the event horizon, and the perturbed equation for $B$ becomes
\be\label{bvv}
\delta B_{,vv} = -\frac{r_h^2}{\kappa_h^2M_p^2 v^2 }\phi^{\prime 2}
\ee
which is solved by
\be\label{deltabsol}
\delta B = -   \frac{ r_h^2 }{ 3\gamma \kappa_h M_p^2 } v
\int_v^\infty \left ( W[\phi(v')] - W[\phi(0)] \right) \frac{dv^\prime }{v^{\prime 2} }
\ee
We have set a constant of integration equal to zero that would have lead to an 
increase in the area that grows linearly
in $v$, rather than a constant value as is consistent with asymptotically dS boundary conditions.
As $v\to0$, we get $\delta B \to 0$, but as
$v\to\infty$ (or as we go to future infinity),
\be
\delta B =  \frac {r_h^2}{3\gamma \kappa_h M_p^2} (W_i - W_f )
\ee
This means that the change in black hole horizon
radius is
\be
\delta r_h =\delta \sqrt{B(r_h)} = \frac{\delta B}{2r_h} 
= \frac{r_h}{ 6 \gamma \kappa_h M_p^2} (W_i - W_f )
\label{deltarh}
\ee
One sees that $\delta B_{,v}\neq 0$ at $v=0$, which is equivalent to the observation
that the event horizon begins to move out before matter has crossed it, which 
follows from the teleological nature of its definition.
The corresponding change in the area of the black hole horizon is
\be\label{changebha}
\delta A_h = \frac{A_h }{3\gamma \kappa_h M_p^2 } (W _i - W_f )
\ee
The area of the black hole horizon increases, as is expected due to accretion 
of the scalar field. The fractional rate of area increase is less for the black hole 
than for the cosmological horizon, since \eqref{changecha} and \eqref{changebha} imply
\be\label{arearatio}
\frac{  \left(\delta A_h  / A_h  \right) }{\left(  \delta A_c/A_c   \right)  } 
= {| \kappa _ c | \over \kappa_h} \  < 1 \  
\ee

%

Now that we have the changes in horizon radii, as a check we can compute 
the change in $\Lambda$ that would be required by the SdS relation 
equation \eqref{sdsparam}, using the changes in horizon radius computed by
integrating the horizon radii, equations \eqref{deltarc} and \eqref{deltarh}
\be
\beal\label{checklambda2}
\delta \Lambda &= -3 H^4 \left ( (2r_c+r_h) \delta r_c 
+ (2r_h+r_c) \delta r_h \right)\\
&=  \frac{H^4 (W_f-W_i)}{\gamma M_p^2}
\left ( (2r_c+r_h) \frac{r_c}{2 \kappa_c }
- (2r_h+r_c) \frac{r_h }{2 \kappa_h} \right)\\
&=  \frac{H^4 \delta W}{\gamma M_p^2}
\left ( \frac{r_c^2}{H^2(r_c-r_h)} - \frac{r_h^2 }{H^2(r_c-r_h)} \right)
= \frac{\delta W}{M_p^2}
\eeal
\ee
as required.

\section{Dynamical thermodynamics}
\label{sec:dynTD}

In this section we explore a number of aspects of the thermodynamics of the 
slow roll evolution of black holes between initial and final Schwarzschild-de 
Sitter states that we have established in the last section.

\subsection{Analysis of horizon growth}

Let us further analyze the results for evolution of the black hole and cosmological 
horizons found in the last section. Using the definition (\ref{gamma})
of $\gamma$ as well as the thermodynamic volume and total horizon area, 
both equations \eqref{changecha} and \eqref{changebha} can be written 
in terms of thermodynamic quantities as
\be\label{deltaageom}
\delta A_\alpha ={\vcal \over 2\pi T_\alpha } {A_\alpha\over  A_{tot} } 
(\Lambda _i - \Lambda _f )  \ , \quad \alpha=h,c
\ee
where $2\pi T_\alpha =|\kappa_\alpha |$ are the horizon temperatures and $\Lambda_{i,f} = W_{i,f}/M_p^2$ are the initial and final values of the cosmological constant.
For a fixed potential $W$, and hence fixed values for $\Lambda_{i,f}$, how much does a 
black hole with initial radius $r_h$ grow? 
Using the formulae for the temperatures and areas in 
Schwarzschild-de Sitter spacetime \eqref{sdsparam}, 
we can express the change in the areas 
in terms of $r_h$, $\Lambda_i$, and $\delta\Lambda=\Lambda_i-\Lambda_f$. For the black hole horizon, one finds that
\be\label{deltaastep}
\delta A_h = A_h { |\delta \Lambda | \over  \Lambda_i} \  
{ 2 r_h ( r_c ^2 + r_h^2 +r_c r_h ) \over ( 2r_h + r_c)(r_h^2 +  r_c^2 )}
\ee
while the corresponding expression for $ \delta A_c$ is obtained by interchanging $r_h$ and $r_c$. 
However, this is not quite what we want, since $r_c$ is still dependent on $r_h$ and $\Lambda_i$. 
This dependence can be dealt with exactly using (\ref{sdsparam}), but 
it is most useful to focus on the limits where the black hole horizon is either small or comparable in size to the cosmological horizon.
One finds that in these limits, the change in the black hole horizon area is given by 
\be\label{bhareachange}
\beal
\delta A_h &\simeq   2 A_h { |\delta \Lambda | \over  \sqrt{3}\Lambda_i}  \   
\times (r_h \sqrt{\Lambda_i} )  \ , \quad r_h \sqrt{\Lambda_i} \ll 1  \\
&\simeq    A_h { |\delta \Lambda | \over  \Lambda_i} \ \  ,   
\quad\quad \quad\quad\quad    r_h \sqrt{\Lambda_i} \sim 1
\eeal
\ee
We see that the fractional growth in area, $\delta A_h/A_h$, is parametrically 
suppressed for small black holes, 
while it is of order $| \delta \Lambda| /\Lambda_i $ for large ones.
Likewise, one can ask how much the cosmological horizon is ``pulled back" 
by the black hole, compared to the case with no black hole. In the limiting cases of small and large black holes, one finds that the change in the cosmological horizon area is given by
\be\label{careachange}
\beal
\delta A_c &\simeq  12 \pi { |\delta \Lambda | \over  \Lambda_i^2}  \  \ , 
\quad r_h \sqrt{\Lambda_i} \ll 1  \\
&\simeq   4 \pi { |\delta \Lambda | \over  \Lambda_i^2}  \ \  ,  
\quad\quad   r_h \sqrt{\Lambda_i} \sim 1
\eeal
\ee
For small black holes, the effect of the black hole on the cosmological horizon growth is negligible.  While for large black holes, it can be reduced by as much as $2/3$.
For a black hole with initial area $1/100$ of the cosmological
horizon area, one finds that the diminution effect is a factor of $1/10$.

\subsection{Two first laws}

Two independent first laws can be derived for asymptotically de Sitter black hole spacetimes 
can be derived \cite{Dolan:2013ft}.   One relates the change in 
area of the black hole horizon to the change in mass, while the other 
relates the change in area of the cosmological horizon to the change in mass.
Including the possibility of a change in the cosmological constant, 
each of these laws has an additional term proportional 
to a thermodynamic volume times $\delta \Lambda$, {\it i.e.}\ a term of the form
$\vcal_\alpha \delta \Lambda$, where $\alpha= h,c$.  One can take the 
difference of the two first laws, such that the mass term drops out
giving
\be\label{firstbetween}
T_h \delta S_h + T_c \delta S_c = \vcal\delta P
\ee
where $\vcal$ is the thermodynamic volume between the black hole and 
cosmological horizons, which was introduced 
above, and we have set
$|\kappa _I | \delta A_I= 8\pi T_I \delta S_I $ and $\Lambda = - 8\pi P$. 
 
Here we are studying the evolution from one Schwarzschild-de Sitter spacetime to another, where 
the change is effected by the rolling scalar field. Combining the changes in the 
horizon areas computed in the previous sections, and given in equations \eqref{changecha}, 
and \eqref{changebha}, one can check that \eqref{firstbetween} is indeed 
satisfied for these evolutions.
This requires use of \eqref{gamma}, which implies that $(A_h +A_c )/(3 \gamma ) = \vcal$. 
This result is interesting because a dynamical scalar
field is beyond the scope of applicability of the derivation in \cite{Dolan:2013ft}, 
and yet our results for a black hole in slow-roll inflation
are still found to satisfy the first law, applied to the differences between the
initial and final de Sitter phases. This agreement suggests that the first law 
might be satisfied continuously along the evolution,
an idea that we return to in the discussion.

\subsection{Temperature and mass for evolving black holes}

A definition of temperature for dynamical black holes was discussed in 
\cite{Hayward:2008jq}, which proposes that a generalized surface gravity is
\be\label{genkappa}
2 \kappa_{dyn} =- \star d \star d \sqrt{B}
\ee
where the Hodge dual $\star$ refers to the 2D spacetime
perpendicular to $\theta$ and $\phi$, or the $U-V$ part, 
and the right hand side is evaluated on the horizon\footnote{ This 
formula differs by a minus sign \eqref{genkappa} from that in \cite{Hayward:2008jq}
due to using different signatures.}.
On the $u-v$ subspace, one has
\be
\star d u = du \qquad , \qquad \star d v = -dv \qquad , \qquad
\star d u \wedge dv = - (g_{uv}^{-1})
\ee
Evaluating (\ref{genkappa}) on the black hole horizon at $u=0$, one finds that
\be\label{tempone}
\beal
2\kappa_{dyn} = -\star d \star d \sqrt{B} &= 
-\star d \left [ \frac{B_{,u} du - B_{,v} dv}{2\sqrt{B}} \right] = 
\left [- \frac{B_{,uv}}{2} + \frac{B_{,u} B_{,v}}{4B} \right] \frac{e^{-2\nu}}{\sqrt{B_0}}\\
&=  B^{-1/2} - \frac{W(\phi)}{M_p^2} \sqrt{B} + \frac{B_{,u} B_{,v} }{4B} e^{-2\nu}B_0^{-1/2}
\eeal
\ee
where the Einstein equation \eqref{Uint} has been used in obtaining the second line. 
One might guess that for slow-roll evolution the black hole temperature would 
instantaneously be that of a Schwarzschild-de Sitter spacetime with the value of 
$\Lambda $ and $r_h$ at that time, and this turns out to be almost the case. 
Using the function $N(r)$ in \eqref{sdsn}, the black hole temperature in Schwarzschild-de 
Sitter is $4\pi T_{sds} ={1\over r_h} - \Lambda r_h $.
For the evolving spacetime, this gives the temperatures in the initial and final states, 
where $r_h$ and $\Lambda $ taking their initial and final values.
Defining a {\it quasi-static temperature} that interpolates between the initial and final 
values along a sequence of Schwarzschild-de Sitter  spacetimes as
\be\label{sdstemp}
4\pi T_{qs} (v) ={1\over r_h (v) } - \frac{W(v)}{M_p^2} r_h (v)
\ee
then this matches the first two terms in \eqref{tempone}. We can evaluate the last term 
in \eqref{tempone} perturbatively. In the background solution, one has $B_{,v}=0$ and
$B_{,u} e^{-2\nu} B_0^{-1/2} = - 4 \kappa_h v $, so that
\be
{B_{,u} B_{,v} \over 4B} e^{-2\nu} B_0^{-1/2} \simeq - { \kappa_h v \over r_h ^2 } \delta B_{,v}
\ee
So the dynamical temperature is given by 
\be\label{temptwo}
2\pi T_{dyn} (v)   = 2\pi T_{qs} (v) - { \kappa_h v \over r_h ^2 } \delta B_{,v}  
\ee
where the derivative of $\delta B$, which follows from \eqref{bvv} 
(or from  \eqref{deltabsol}), is given by 
\be\label{bv}
\delta B_{,v} = -  \frac{r_h^2 }{ 3\gamma \kappa_h M_p^2} 
\left[  \int_v^\infty  \delta W[\phi(v')]  \frac{dv^\prime }{v^{\prime 2} } 
- \frac{\delta W[\phi (v)] }{v}  \right]
\ee

Now consider the late time behaviour of this temperature, when the integral in 
$\delta B$ is approximately given by\footnote{For
more detailed discussion of the asymptotic behaviour of $B$, see \S 
\ref{sec:example}, where the dynamics of the black hole system is 
computed in detail for a sample potential.}
 $\int \delta W/v^2 \simeq \delta W/v$.  Substituting this in to $\delta B$ and \eqref{bv}, at late times
the dynamical temperature is given by the quasi-static approximation, 
which expanded to first order is
\be\label{latetemp}
T_{dyn}  \simeq T_{qs} \simeq T_{sds, i} 
- \frac{(W_i - W[\phi (v) ])}{24\pi\gamma \kappa_h M_p^2 } 
\left (1+ \frac{W_i r_h^2}{M_p^2 } - 6 \gamma \kappa_h r_h^2 \right ) 
\ee
where $T_{sds, i} $ is the temperature of the initial SdS spacetime.

A definition of the dynamical mass  can be found by considering the first law
relating the change in mass to the changes in black hole area and $\Lambda$.
Perturbations about a static black hole with positive $\Lambda$ satisfy \cite{Dolan:2013ft} 
\be\label{otherfirst}
\delta M = 
T_h \delta S_h - {\cal V}_h \delta P
\ee
where $ {\cal V}_h= 4\pi r_h^3 /3$ is the black hole thermodynamic volume, here
given for SdS spacetime. As noted in the context of the first law 
formulated between the two horizons \eqref{firstbetween}, a dynamical cosmological 
constant due to 
a scalar field potential is beyond the scope of the derivation 
of \eqref{otherfirst}. However, since we found that 
\eqref{firstbetween} is true for our solutions, let us assume that \eqref{otherfirst}
also holds as well and see what it implies for the mass. This is equivalent to assuming that at late times, when the 
stress-energy is again dominated by the smaller
effective $\Lambda_f$, that the metric can be put into static SdS form with the 
evolved values of $r_h$ and $r_c$. The final mass $M_f$ is then given by the SdS relations 
\eqref{sdsparam}. Explicitly, substituting $\delta A_h$ from \eqref{changebha} 
and $\delta \Lambda = (W_f -W_i )/M_p^2$ into \eqref{otherfirst} gives
\be\label{deltam}
\delta M 
= {r_c r_h \over r_c^2 +r_h^2 }M { |\delta \Lambda | \over \Lambda}
\ee
Analogous to the check we did on the change in $\Lambda$ \eqref{checklambda2}, 
one can vary $M$ directly from \eqref{sdsparam},
substitute in our results for the changes in $r_h ,\   r_c$ and $\Lambda$, and find 
the same answer as in \eqref{deltam}.

Lastly, it is interesting to assemble the terms on the right hand side of 
\eqref{otherfirst} as follows. We have that
\be
\beal
T_{dyn} \delta S &= \frac{1}{4G} \left [  B^{-1/2}-\frac{W(\phi)}{M_p^2} \sqrt{B} 
- \frac{v\kappa_h \delta B_{,v}}{B}
\right] \delta B\\
&= \frac{1}{4G} \left [B^{-1/2}-\frac{W(\phi)}{M_p^2} \sqrt{B} \right] \delta B + {\cal{O}}
(\delta B)^2
\eeal
\ee
Meanwhile, 
\be
V_h \delta P = \frac{4\pi B^{3/2}}{3} \delta (-W)
\ee
so
\be
\beal
T_{dyn} \delta S_h + V_h\delta P &=  \frac{\delta B}{4G\sqrt{B}}- 2\pi W \sqrt{B} \delta B 
- \frac{4\pi}{3} B^{3/2} \delta W\\
&= \frac{1}{2G} \delta \left [ \sqrt{B} - \frac{W}{3M_p^2} B^{3/2} \right]\Bigg |_{u=0}
\eeal
\ee
This quantity is defined on the black hole horizon, and if our assumptions are 
correct, is equal to $\delta M$ in \eqref{deltam}.
It also suggests that the mass is given by
\be
M = \frac{1}{2G} \left [ \sqrt{B} - \frac{W}{3M_p^2} B^{3/2} \right] \Big |_{u=0}
\ee
In the static case, where
$B = r^2$, this is precisely the definition of $M$. 
Supporting this interpretation is that integrating \eqref{Vint}, gives that to leading 
order the additional piece in the dynamical temperature is
\be
\delta B_{,v} = -\frac{1}{M_p^2} \int B \phi_{,v}^2 = -\frac{1}{M_p^2} \int T_{vv}
\ee
that is, the `mass' contribution due to scalar accretion onto the black hole. 
In general, to complete the argument, 
we would want to show that the quantity evaluated at $u=0$ is equal to 
a quantity defined on future spacelike infinity.

\section{Illustrative example}
\label{sec:example}

In this paper, we have derived general results for the accreting black
hole. It is helpful to illustrate these with a test-case example, using our
standard double well potential \eqref{classicv}, which will  allow us to explore the effects
of varying the black hole mass and slow roll parameters.
Recall that the solution to the slow-roll equation, $3\gamma\phi' = - W'(\phi)$,
for \eqref{classicv} is
\be
\phi^2 = \eta^2 \frac{e^{H_i^2 \Gamma x/2\gamma}}{e^{H_i^2 \Gamma x/2\gamma}+1}
\ee
We start by focussing on the black hole event horizon where 
$\kappa_h x \simeq \log (2\kappa_h v)$, and
compute the dynamical horizon area and temperature as a function of
${\hat v} = 2\kappa_h v$. Writing $a=\Gamma H_i^2/2\gamma\kappa_h$, we have
\be
\delta W = W[\phi] - W[0] = -\frac{3 H_i^2 \Gamma\eta^2}{16} 
\frac{\hat{v}^a (2+\hat{v}^a)}
{(\hat{v}^a+1)^2}
\ee
and hence \eqref{deltabsol} gives
\be
{\cal A} = 4\pi B = {\cal A}_0 \left ( 1 - \frac{a\Delta {\hat v}}{8}
{\cal I}[\hat{v},a] \right)
\ee
Here $\Delta = \eta^2/M_p^2$ represents the strength of the gravitational 
interaction of the scalar field, and
\be
\beal
{\cal I}[ {\hat v},a] &= - \int_{\hat v}^\infty \frac{y^a(2+y^a) dy}{y^2(1+y^a)^2}\\
&= - \frac{(1+a(1+ \hat{v}^a))}{a \hat{v} (1+ \hat{v}^a)}
+ \frac{1+a}{a \hat{v}} \left ( 1 - ~_2F_1 \left [ 1, \frac1a;\frac{1+a}{a}; -\hat{v}^{-a} \right]\right)
\eeal
\ee
is a dimensionless integral. 
Meanwhile, the dynamical temperature is found to be
\be
T_{dyn} = T_{init} + \frac{a\Delta}{32\pi} \left [
\kappa_h (3\gamma r_h +1)  \frac{\hat{v}^{a}(2+\hat{v}^{a})}{(1+\hat{v}^{a})^2}
+ \frac{{\hat v}}{r_h} {\cal I}[\hat{v},a] \right ]
\ee
\begin{figure}
\centering
\includegraphics[scale=0.6]{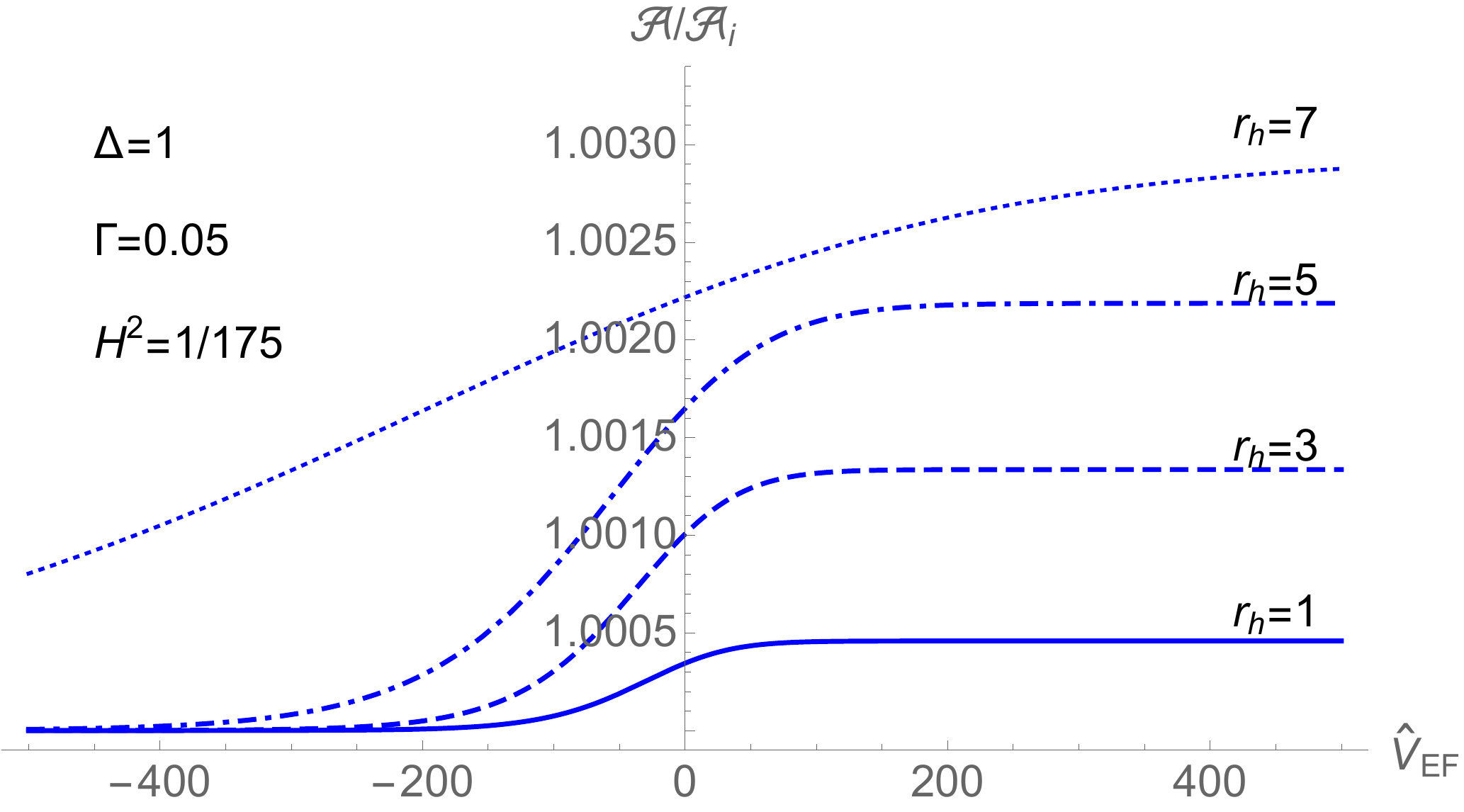}
\caption{Illustration of evolution of the horizon area (${\cal A}_h/{\cal A}_i$)
as a function of the Eddington-Finkelstein advanced time coordinate,
scaled by the Hubble parameter, on the black hole horizon.  
We see that for a larger initial black hole, the increase in black hole 
horizon area is both larger and more gradual, than for small black holes.}
\label{fig:dynamicalhorizon}
\end{figure}

Having extracted the dimensionful dependence, we can now see how
the various parameters impact the evolution of the scalar and
hence the black hole horizon. First, it is clear that the overall gravitational
strength of the scalar, $\Delta$, simply scales the overall 
magnitude of the variation of the area and temperature. The parameter
$a$ on the other hand not only scales the magnitude of these variations, 
but also their rapidity, as would be expected since $a$ is directly 
dependent on the rate of slow-roll of the scalar field. 
Given the expressions for $\gamma$ and $\kappa_h$, we see that
\be
a \sim \frac{2\Gamma H_i^2}{\gamma \kappa_h}
= \frac{\Gamma (r_c^2/r_h^2+ r_c/r_h +1)}
{(r_c^2/r_h^2+1)(r_c/r_h +2)} < \frac\Gamma2
\ee

\begin{figure}
\centering
\includegraphics[scale=0.6]{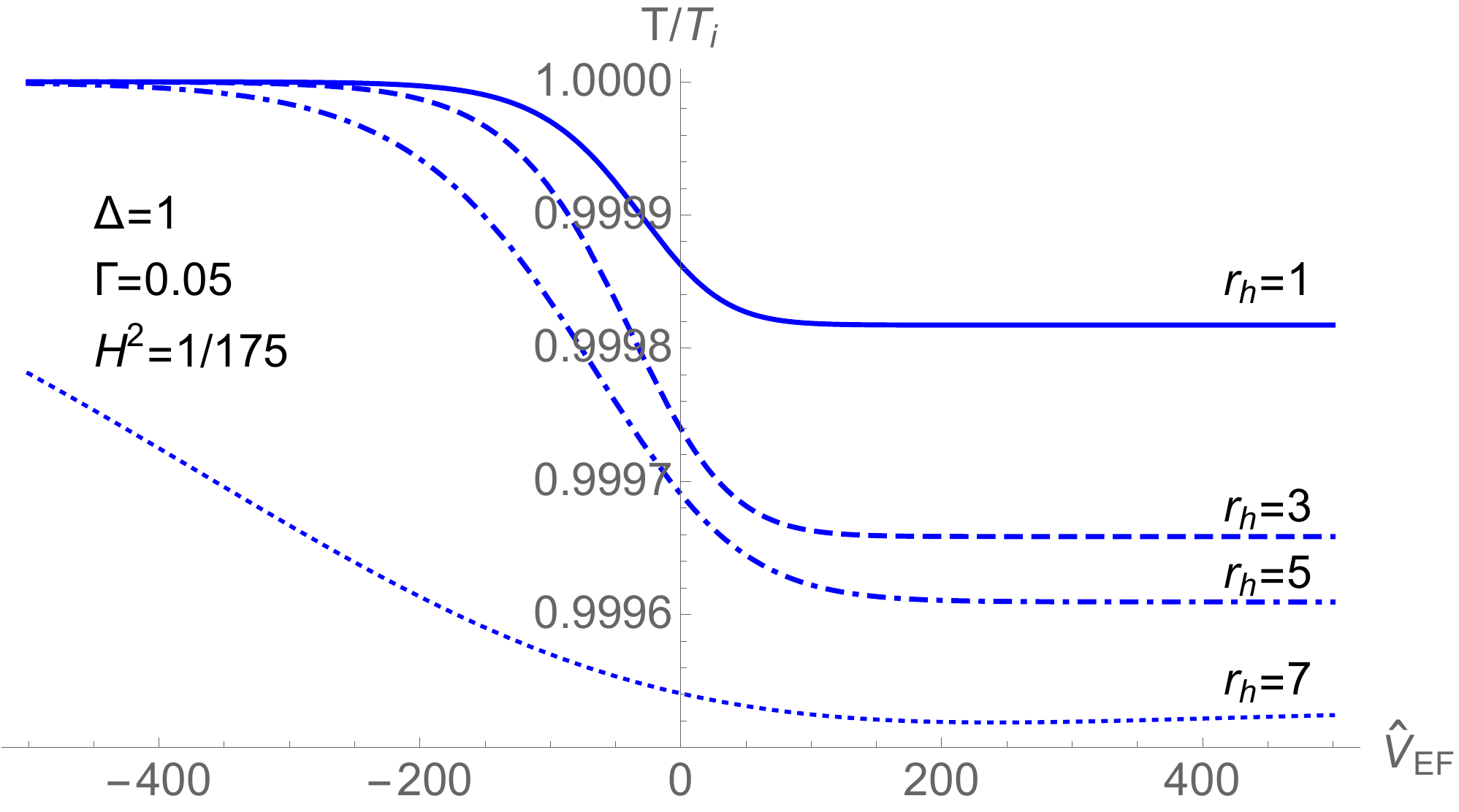}
\caption{Illustration of evolution of 
temperature ($T/T_i$) as a function of the Eddington-Finkelstein advanced time
coordinate on the black hole horizon.  We see that for larger black holes the decrease in black hole horizon temperature is both larger and more gradual than for small black holes.}
\label{fig:dynamicaltemp}
\end{figure}

It is perhaps most useful to display the variations of the black hole variables
as a function of the local (normalised) Eddington-Finkelstein 
advanced time co-ordinate on the event horizon,
$\hat{V}_{EF} = H(t+r^\star) = \frac{H}{\kappa_h}\log[2\kappa_h v] \sim H x$.
Figure \ref{fig:dynamicalhorizon} shows a plot of the variation of horizon
area with advanced time, while Figure \ref{fig:dynamicaltemp} shows the variation of 
temperature. In both cases, rather large values of $\Delta$ and $\Gamma$
have been chosen to emphasize the effects. The differing gradations of
``slow-roll'' on display arise because of the differing black hole : cosmological
horizon ratios -- recall that the true slow-roll parameter $\gamma$, given in
\eqref{gamma}, for the scalar in the presence of the black hole has a strong
dependence on the geometry. 

\section{Concluding remarks}

Central concepts in gravitational thermodynamics are horizon areas and 
temperatures, and the relations between them. The analysis presented 
here allows the study of these quantities in a ``mildly" 
dynamical setting, namely the evolution from 
one SdS spacetime to a second SdS with a smaller cosmological constant, 
in a perturbative and slow-roll approximation.
An advantage of these boundary conditions is that  the initial and final states are equilibria
with approximate Killing horizons and associated temperatures. 
Within these approximations we have solved the
Einstein plus scalar field system to extract the growth of the black hole 
and cosmological horizons for a general scalar potential that 
has a maximum and a minimum. The results are expressed in terms of the
change in the effective cosmological constant, as dictated by the potential, 
and geometrical properties of the initial spacetime. Using a proposed definition of
dynamical temperature, the temperatures of each horizon are found to 
decrease between the initial and final values as the horizons grow.

One of the interesting features of the solutions is that the first law of thermodynamics
\eqref{firstbetween}, formulated between the two horizons, holds between the 
initial and final SdS states. This brings up two questions
for further study. One is to derive the first law including the stress-energy of 
a scalar field, which generates additional contributions at
the black hole and cosmological horizons, and verify that  these contributions 
vanish for our solutions. Second,
this result suggests that the first law might be satisfied not only between the early 
and late time SdS metrics, but continuously
along the evolution. More strongly, is there a solution for the metric functions 
which illustrates that the metric 
is quasi-SdS at each time? In general, it would be advantageous to have 
explicit expressions for the full metric across the range of $r$ and $x$.
This would facilitate analyzing the flow of energy-momentum throughout the 
volume, to test the definition of dynamical temperature
by studying the near-horizon metric, and to follow the evolution of the 
mass-like quantity that interpolates between the initial and final mass parameters.
Another interesting direction for further work is to transform the analysis to 
cosmological coordinates, which 
are likely more convenient for ascertaining how a black hole affects the 
important predictions of inflation, such as the spectrum of 
quantum perturbations and reheating.
 
\section*{Acknowledgements} 

The authors would like to thank Kostas Skenderis for helpful conversations.
RG is supported in part by STFC (Consolidated Grant ST/J000407/1).
RG also acknowledges support from the Wolfson Foundation and Royal Society, and
Perimeter Institute for Theoretical Physics. 
Research at Perimeter Institute is supported by the Government of
Canada through the Department of Innovation, Science and Economic 
Development Canada and by the Province of Ontario through the
Ministry of Research, Innovation and Science.

\appendix
\section{Cosmological tension}\label{admsection}

The ADM cosmological tension charges for an asymptotically future de Sitter 
spacetime were constructed in \cite{Kastor:2016bnm}.
Generally, an ADM charge corresponding to an asymptotic symmetry of a 
spacetime is defined via Hamiltonian perturbation theory 
\cite{Regge:1974zd}. We follow the general prescription presented 
in \cite{Traschen:2001pb}, which we briefly outline here.
The construction starts with foliating the spacetime by $(D-1)$-dimensional 
hypersurfaces $\Sigma$ with unit normal $n^a$ such that the metric can be decomposed as 
\begin{align*}
g_{ab}= (n\cdot n) n_an_b+s_{ab}
\end{align*}
where $n_a n^a =\pm 1$ and $s_{ab}$ denotes the induced metric on the slice(s), 
satisfying the orthogonality relation $s_{ab}n^b=0$.
Let $(s_{ab},\pi^{ab})$ denote the Hamiltonian initial data on a slice $\Sigma$ 
and $(h_{ab},p^{ab})$ be perturbations linearized about the background denoted 
by $(\bar{s}_{ab},\bar{\pi}^{ab})$. Furthermore, let $\xi^a$ be a Killing vector of the 
background, which we project along the slice $\Sigma$ and its normal 
according to $\xi^a=Fn^a+\beta^a$ such that $n_a\beta^a=0$.

The ADM charge corresponding to the Killing vector $\xi^a$ is then 
defined by an integral over a $(D-2)$-dimensional boundary at infinity on $\Sigma$ given by
\begin{align}{\label{admcharge}}
Q(\xi)=-\dfrac{1}{16\pi G}\int_{\partial\Sigma_{\infty}}da_cB^c
\end{align}
where
\begin{align}\label{admbt}
B^a= &F(\bar{D}^ah-\bar{D}_bh^{ab})-h\bar{D}^aF+h^{ab}\bar{D}_bF \\ &\nonumber
+\dfrac{1}{\sqrt{s}}\beta^b(\bar{\pi}^{cd}h_{cd}\bar{s}^a{}_b-2\bar{\pi}^{ac}h_{bc}-2p^a{}_b)
\end{align}
and $\bar{D}_a$ is the covariant derivative operator compatible with the 
background metric $\bar{s}_{ab}$. 
Note that to simply define the charge $Q$ one only needs the symmetry 
(and the foliation) asymptotically.
However, if the symmetry holds througout the spacetime, this set-up can then be used to 
prove a first law \cite{Traschen:2001pb}.

The ADM mass results when a spacetime has an asymptotic static 
Killing field  at spatial infinity and choosing $\Sigma$ to be a timelike slice with 
unit timelike normal, $n_a n^a =-1$.
On the other hand, the construction can be used for a cosmological 
spacetime with an asymptotic spatial translation Killing field $\xi$ and taking 
$\Sigma$ to have a unit spacelike normal, $n_a n^a =1$.
Then the boundary $\partial\Sigma$ is in the asymptotic future. 
The resulting ADM charge is a cosmological tension. 

If the spacetime is anisotropic but homogeneous, as was considered in 
\cite{Kastor:2016bnm}, then the tensions are distinct.
The inflationary spacetimes studied in this paper are isotropic and homogeneous,
so there are three equal tensions.  Using the formulae derived in \cite{Kastor:2016bnm} 
to process the boundary term \eqref{admbt} for the underdamped asymptotically de Sitter  case,
and averaging over a period of oscillation, gives the cosmological tension 
\eqref{tension}. In the overdamped
and critically damped  cases, the metric functions decay too slowly
to balance the growth of the volume element, and the tension diverges.

\providecommand{\href}[2]{#2}
\begingroup\raggedright\endgroup  

\begin{thebibliography}{10}

\bibitem{Carr:2009jm}  
B.~J.~Carr, K.~Kohri, Y.~Sendouda and J.~Yokoyama,
{\it New cosmological constraints on primordial black holes},
Phys.\ Rev.\ D {\bf 81}, 104019 (2010)
\href{http://xxx.lanl.gov/abs/0912.5297}{{\tt arXiv:0912.5297 [astro-ph.CO]}}.
  
\bibitem{Carr:2016drx} 
B.~Carr, F.~Kuhnel and M.~Sandstad,
{\it Primordial Black Holes as Dark Matter},
Phys.\ Rev.\ D {\bf 94}, no. 8, 083504 (2016)
\href{http://xxx.lanl.gov/abs/1607.06077}{{\tt arXiv:1607.06077 [astro-ph.CO]}}.

\bibitem{Sherkatghanad:2015rga} 
Z.~Sherkatghanad and R.~H.~Brandenberger,
{\it The Effect of Primordial Non-Gaussianities on the Seeds of Super-Massive Black Holes},
\href{http://xxx.lanl.gov/abs/1508.00968}{{\tt arXiv:1508.00968 [astro-ph.CO]}}.

\bibitem{Bird:2016dcv} 
S.~Bird, I.~Cholis, J.~B.~Mu–oz, Y.~Ali-Ha•moud, M.~Kamionkowski, 
E.~D.~Kovetz, A.~Raccanelli and A.~G.~Riess,
{\it Did LIGO detect dark matter?},
Phys.\ Rev.\ Lett.\  {\bf 116}, no. 20, 201301 (2016)
\href{http://xxx.lanl.gov/abs/1603.00464}{{\tt arXiv:1603.00464 [astro-ph.CO]}}.

\bibitem{Kawasaki:2016pql}
M.~Kawasaki, A.~Kusenko, Y.~Tada and T.~T.~Yanagida,
{\it Primordial black holes as dark matter in supergravity inflation models},
Phys.\ Rev.\ D {\bf 94} (2016) no.8,  083523
\href{http://xxx.lanl.gov/abs/1606.07631}{{\tt arXiv:1606.07631 [astro-ph.CO]}}.
 
\bibitem{Bean:2002kx} 
R.~Bean and J.~Magueijo,
{\it Could supermassive black holes be quintessential primordial black holes?},
Phys.\ Rev.\ D {\bf 66}, 063505 (2002)
\href{http://xxx.lanl.gov/abs/astro-ph/0204486}{{\tt astro-ph/0204486}}.
  
\bibitem{Kannike:2017bxn}
K.~Kannike, L.~Marzola, M.~Raidal and H.~VeermŠe,
{\it Single Field Double Inflation and Primordial Black Holes},
\href{http://xxx.lanl.gov/abs/1705.06225}{{\tt arXiv:1705.06225 [astro-ph.CO]}}.
  
\bibitem{Domcke:2017fix} 
V.~Domcke, F.~Muia, M.~Pieroni and L.~T.~Witkowski,
{\it PBH dark matter from axion inflation},
\href{http://xxx.lanl.gov/abs/1704.03464}{{\tt arXiv:1704.03464 [astro-ph.CO]}}.

\bibitem{Georg:2017mqk}
J.~Georg and S.~Watson,
{\it A Preferred Mass Range for Primordial Black Hole Formation and 
Black Holes as Dark Matter Revisited},
\href{http://xxx.lanl.gov/abs/1703.04825}{{\tt arXiv:1703.04825 [astro-ph.CO]}}.

\bibitem{Mediavilla:2017bok} 
E.~Mediavilla, J.~JimŽnez-Vicente, J.~A.~Mu–oz, H.~Vives-Arias and J.~Calder—n-Infante,
{\it Limits on the Mass and Abundance of Primordial Black Holes from 
Quasar Gravitational Microlensing},
Astrophys.\ J.\  {\bf 836}, no. 2, L18 (2017)
\href{http://xxx.lanl.gov/abs/1702.00947}{{\tt arXiv:1702.00947 [astro-ph.GA]}}.

\bibitem{Kastor:2009wy} 
D.~Kastor, S.~Ray and J.~Traschen,
{\it Enthalpy and the Mechanics of AdS Black Holes},
Class.\ Quant.\ Grav.\  {\bf 26}, 195011 (2009)
\href{http://xxx.lanl.gov/abs/0904.2765}{{\tt arXiv:0904.2765 [hep-th]}}.

\bibitem{Dolan:2013ft} 
B.~P.~Dolan, D.~Kastor, D.~Kubiznak, R.~B.~Mann and J.~Traschen,
{\it Thermodynamic Volumes and Isoperimetric Inequalities for de Sitter Black Holes},
Phys.\ Rev.\ D {\bf 87}, no. 10, 104017 (2013)
\href{http://xxx.lanl.gov/abs/1301.5926}{{\tt arXiv:1301.5926 [hep-th]}}.

\bibitem{Kastor:2016bnm} 
D.~Kastor, S.~Ray and J.~Traschen,
{\it Genuine Cosmic Hair},
Class.\ Quant.\ Grav.\  {\bf 34}, no. 4, 045003 (2017)
\href{http://xxx.lanl.gov/abs/1608.04641}{{\tt arXiv:1608.04641 [hep-th]}}.

\bibitem{Sudarsky:2002mk} 
D.~Sudarsky and J.~A.~Gonzalez,
{\it On black hole scalar hair in asymptotically anti-de Sitter space-times},
Phys.\ Rev.\ D {\bf 67}, 024038 (2003)
\href{http://xxx.lanl.gov/abs/gr-qc/0207069}{{\tt arXiv:gr-qc/0207069}}.

\bibitem{Henneaux:2002wm} 
M.~Henneaux, C.~Martinez, R.~Troncoso and J.~Zanelli,
{\it Black holes and asymptotics of 2+1 gravity coupled to a scalar field},
Phys.\ Rev.\ D {\bf 65}, 104007 (2002)
\href{http://xxx.lanl.gov/abs/hep-th/0201170}{{\tt arXiv:hep-th/0201170}}.

\bibitem{Hertog:2004dr} 
T.~Hertog and K.~Maeda,
{\it Black holes with scalar hair and asymptotics in N = 8 supergravity},
JHEP {\bf 0407}, 051 (2004)
\href{http://xxx.lanl.gov/abs/hep-th/0404261}{{\tt arXiv:hep-th/0404261}}.

\bibitem{Martinez:2004nb} 
C.~Martinez, R.~Troncoso and J.~Zanelli,
{\it Exact black hole solution with a minimally coupled scalar field},
Phys.\ Rev.\ D {\bf 70}, 084035 (2004)
\href{http://xxx.lanl.gov/abs/hep-th/0406111}{{\tt arXiv:hep-th/0406111}}.

\bibitem{Radu:2004xp} 
E.~Radu and D.~H.~Tchrakian,
{\it New hairy black hole solutions with a dilaton potential},
Class.\ Quant.\ Grav.\  {\bf 22}, 879 (2005)
\href{http://xxx.lanl.gov/abs/hep-th/0410154}{{\tt arXiv:hep-th/0410154}}.

\bibitem{Winstanley:2005fu} 
E.~Winstanley,
{\it Dressing a black hole with non-minimally coupled scalar field hair},
Class.\ Quant.\ Grav.\  {\bf 22}, 2233 (2005)
\href{http://xxx.lanl.gov/abs/gr-qc/0501096}{{\tt arXiv:gr-qc/0501096}}.
 
\bibitem{Faedo:2015jqa} 
F.~Faedo, D.~Klemm and M.~Nozawa,
{\it Hairy black holes in $\rm{N} = 2$ gauged supergravity},
JHEP {\bf 1511}, 045 (2015)
\href{http://xxx.lanl.gov/abs/1505.02986}{{\tt arXiv:1505.02986 [hep-th]}}.

\bibitem{Anabalon:2015xvl}
A.~Anabalon, D.~Astefanesei, D.~Choque and C.~Martinez,
{\it Trace Anomaly and Counterterms in Designer Gravity},
JHEP {\bf 1603} (2016) 117
\href{http://xxx.lanl.gov/abs/1511.08759}{{\tt arXiv:1511.08759 [hep-th]}}.

\bibitem{Hertog:2004ns} 
T.~Hertog and G.~T.~Horowitz,
{\it Designer gravity and field theory effective potentials},
Phys.\ Rev.\ Lett.\  {\bf 94}, 221301 (2005)
\href{http://xxx.lanl.gov/abs/hep-th/0412169}{{\tt arXiv:hep-th/0412169}}.

\bibitem{Henneaux:2006hk} 
M.~Henneaux, C.~Martinez, R.~Troncoso and J.~Zanelli,
{\it Asymptotic behavior and Hamiltonian analysis of anti-de Sitter gravity coupled to scalar fields},
Annals Phys.\  {\bf 322}, 824 (2007)
\href{http://xxx.lanl.gov/abs/hep-th/0603185}{{\tt arXiv:hep-th/0603185}}.

\bibitem{Kaplan:2014dia} 
J.~Kaplan and J.~Wang,
{\it An Effective Theory for Holographic RG Flows},
JHEP {\bf 1502}, 056 (2015)
\href{http://xxx.lanl.gov/abs/1406.4152}{{\tt arXiv:1406.4152 [hep-th]}}.


\bibitem{Skenderis:2006jq} 
K.~Skenderis and P.~K.~Townsend,
{\it Hidden supersymmetry of domain walls and cosmologies},
Phys.\ Rev.\ Lett.\  {\bf 96}, 191301 (2006)
\href{http://xxx.lanl.gov/abs/hep-th/0602260}{{\tt arXiv:hep-th/0602260}}.
 
\bibitem{Bazeia:2007vx} 
D.~Bazeia, F.~A.~Brito and F.~G.~Costa,
{\it First-order framework and domain-wall/brane-cosmology correspondence},
Phys.\ Lett.\ B {\bf 661}, 179 (2008)
\href{http://xxx.lanl.gov/abs/0707.0680}{{\tt arXiv:0707.0680 [hep-th]}}.
 
\bibitem{McFadden:2010na} 
P.~McFadden and K.~Skenderis,
{\it The Holographic Universe},
J.\ Phys.\ Conf.\ Ser.\  {\bf 222}, 012007 (2010)
\href{http://xxx.lanl.gov/abs/1001.2007}{{\tt arXiv:1001.2007 [hep-th]}}.

\bibitem{Bzowski:2012ih} 
A.~Bzowski, P.~McFadden and K.~Skenderis,
{\it Holography for inflation using conformal perturbation theory},
JHEP {\bf 1304}, 047 (2013)
\href{http://xxx.lanl.gov/abs/1211.4550}{{\tt arXiv:1211.4550 [hep-th]}}.

\bibitem{Kol:2013msa} 
U.~Kol,
{\it On the dual flow of slow-roll Inflation},
JHEP {\bf 1401}, 017 (2014)
\href{http://xxx.lanl.gov/abs/1309.7344}{{\tt arXiv:1309.7344 [hep-th]}}.

\bibitem{Chadburn:2013mta} 
S.~Chadburn and R.~Gregory,
{\it Time dependent black holes and scalar hair},
Class.\ Quant.\ Grav.\  {\bf 31}, no. 19, 195006 (2014)
\href{http://xxx.lanl.gov/abs/1304.6287}{{\tt arXiv:1304.6287 [gr-qc]}}.

\bibitem{Hayward:2008jq} 
S.~A.~Hayward, R.~Di Criscienzo, L.~Vanzo, M.~Nadalini and S.~Zerbini,
{\it Local Hawking temperature for dynamical black holes},
Class.\ Quant.\ Grav.\  {\bf 26}, 062001 (2009)
\href{http://xxx.lanl.gov/abs/0806.0014}{{\tt arXiv:0806.0014 [gr-qc]}}.


\bibitem{Girardello:1998pd} 
L.~Girardello, M.~Petrini, M.~Porrati and A.~Zaffaroni,
{\it Novel local CFT and exact results on perturbations of N=4 superYang Mills from AdS dynamics},
JHEP {\bf 9812}, 022 (1998)
\href{http://xxx.lanl.gov/abs/hep-th/9810126}{{\tt arXiv:hep-th/9810126}}.

\bibitem{Freedman:1999gp} 
D.~Z.~Freedman, S.~S.~Gubser, K.~Pilch and N.~P.~Warner,
{\it Renormalization group flows from holography supersymmetry and a c theorem},
Adv.\ Theor.\ Math.\ Phys.\  {\bf 3}, 363 (1999)
\href{http://xxx.lanl.gov/abs/hep-th/9904017}{{\tt arXiv:hep-th/9904017}}.

\bibitem{Skenderis:1999mm} 
K.~Skenderis and P.~K.~Townsend,
{\it Gravitational stability and renormalization group flow},
Phys.\ Lett.\ B {\bf 468}, 46 (1999)
\href{http://xxx.lanl.gov/abs/hep-th/9909070}{{\tt arXiv:hep-th/9909070}}.


\bibitem{Skenderis:2006rr} 
K.~Skenderis and P.~K.~Townsend,
{\it Hamilton-Jacobi method for curved domain walls and cosmologies},
Phys.\ Rev.\ D {\bf 74}, 125008 (2006)
\href{http://xxx.lanl.gov/abs/hep-th/0609056}{{\tt arXiv:hep-th/0609056}}.

\bibitem{Kehagias:2000au} 
A.~Kehagias and K.~Tamvakis,
{\it Localized gravitons, gauge bosons and chiral fermions in smooth 
spaces generated by a bounce},
Phys.\ Lett.\ B {\bf 504}, 38 (2001)
\href{http://xxx.lanl.gov/abs/hep-th/0010112}{{\tt arXiv:hep-th/0010112}}.

\bibitem{Skenderis:2006fb} 
K.~Skenderis and P.~K.~Townsend,
{\it Pseudo-Supersymmetry and the Domain-Wall/Cosmology Correspondence},
J.\ Phys.\ A {\bf 40}, 6733 (2007)
\href{http://xxx.lanl.gov/abs/hep-th/0610253}{{\tt arXiv:hep-th/0610253}}.

\bibitem{Frolov:2002va} 
A.~V.~Frolov and L.~Kofman,
{\it Inflation and de Sitter thermodynamics},
JCAP {\bf 0305}, 009 (2003)
\href{http://xxx.lanl.gov/abs/hep-th/0212327}{{\tt hep-th/0212327}}.

\bibitem{Liddle:1994dx} 
A.~R.~Liddle, P.~Parsons and J.~D.~Barrow,
{\it Formalizing the slow roll approximation in inflation},
Phys.\ Rev.\ D {\bf 50}, 7222 (1994)
\href{http://xxx.lanl.gov/abs/astro-ph/9408015}{{\tt arXiv:astro-ph/9408015}}.

\bibitem{Wald:1983ky} 
R.~M.~Wald,
{\it Asymptotic behavior of homogeneous cosmological models in the 
presence of a positive cosmological constant},
Phys.\ Rev.\ D {\bf 28}, 2118 (1983).

\bibitem{Traschen:2001pb} 
J.~H.~Traschen and D.~Fox,
{\it Tension perturbations of black brane space-times},
Class.\ Quant.\ Grav.\  {\bf 21}, 289 (2004)
\href{http://xxx.lanl.gov/abs/gr-qc/0103106}{{\tt arXiv:gr-qc/0103106}}.
  
\bibitem{El-Menoufi:2013pza} 
B.~Mahmoud El-Menoufi, B.~Ett, D.~Kastor and J.~Traschen,
{\it Gravitational Tension and Thermodynamics of Planar AdS Spacetimes},
Class.\ Quant.\ Grav.\  {\bf 30}, 155003 (2013)
\href{http://xxx.lanl.gov/abs/1302.6980}{{\tt arXiv:1302.6980 [hep-th]}}.

\bibitem{Breitenlohner:1982jf} 
P.~Breitenlohner and D.~Z.~Freedman,
{\it Stability in Gauged Extended Supergravity},
Annals Phys.\  {\bf 144}, 249 (1982).


\bibitem{McVittie:1933zz}
G.~C.~McVittie,
{\it The mass-particle in an expanding universe},
Mon.\ Not.\ Roy.\ Astron.\ Soc.\  {\bf 93}, 325 (1933).



\bibitem{Kastor:1992nn} 
  D.~Kastor and J.~H.~Traschen,
{\it Cosmological multi - black hole solutions},
  Phys.\ Rev.\ D {\bf 47}, 5370 (1993)
\href{http://xxx.lanl.gov/abs/hep-th/9212035}{{\tt  hep-th/9212035}}.

\bibitem{Gibbons:2009dr} 
G.~W.~Gibbons and K.~i.~Maeda,
{\it Black Holes in an Expanding Universe},
Phys.\ Rev.\ Lett.\  {\bf 104}, 131101 (2010)
\href{http://xxx.lanl.gov/abs/0912.2809}{{\tt arXiv:0912.2809 [gr-qc]}}.

\bibitem{Jacobson:1999vr} 
T.~Jacobson,
{\it Primordial black hole evolution in tensor scalar cosmology},
Phys.\ Rev.\ Lett.\  {\bf 83}, 2699 (1999)
\href{http://xxx.lanl.gov/abs/astro-ph/9905303}{{\tt astro-ph/9905303}}.

\bibitem{Saida:2000at} 
H.~Saida and J.~Soda,
{\it Black holes and a scalar field in expanding universe},
Class.\ Quant.\ Grav.\  {\bf 17}, 4967 (2000)
\href{http://xxx.lanl.gov/abs/gr-qc/0006058}{{\tt gr-qc/0006058}}.

\bibitem{Harada:2004pf} 
T.~Harada and B.~J.~Carr,
{\it Growth of primordial black holes in a universe containing a massless scalar field},
Phys.\ Rev.\ D {\bf 71}, 104010 (2005)
\href{http://xxx.lanl.gov/abs/astro-ph/0412135}{{\tt astro-ph/0412135}}.

\bibitem{Sultana:2005tp} 
J.~Sultana and C.~C.~Dyer,
{\it Cosmological black holes: A black hole in the Einstein-de Sitter universe},
Gen.\ Rel.\ Grav.\  {\bf 37}, 1347 (2005).

\bibitem{Faraoni:2007es} 
V.~Faraoni and A.~Jacques,
{\it Cosmological expansion and local physics},
Phys.\ Rev.\ D {\bf 76}, 063510 (2007)
\href{http://xxx.lanl.gov/abs/0707.1350}{{\tt [arXiv:0707.1350 [gr-qc]]}}.

\bibitem{Carrera:2008pi} 
M.~Carrera and D.~Giulini,
{\it On the influence of global cosmological expansion on the 
dynamics and kinematics of local systems},
Rev.\ Mod.\ Phys.\ {\bf 82} 169 (2010)
\href{http://xxx.lanl.gov/abs/0810.2712}{{\tt [arXiv:0810.2712 [gr-qc]]}}.

\bibitem{Rodrigues:2009eg} 
M.~G.~Rodrigues and A.~Saa,
{\it Accretion of nonminimally coupled scalar fields into black holes},
Phys.\ Rev.\ D {\bf 80}, 104018 (2009)
\href{http://xxx.lanl.gov/abs/0909.3033}{{\tt arXiv:0909.3033 [gr-qc]}}.

\bibitem{Carr:2010wk} 
B.~J.~Carr, T.~Harada and H.~Maeda,
{\it Can a primordial black hole or wormhole grow as fast as the universe?},
Class.\ Quant.\ Grav.\  {\bf 27}, 183101 (2010)
\href{http://xxx.lanl.gov/abs/1003.3324}{{\tt arXiv:1003.3324 [gr-qc]}}.

\bibitem{UrenaLopez:2011fd} 
L.~A.~Urena-Lopez and L.~M.~Fernandez,
{\it Black holes and the absorption rate of cosmological scalar fields},
Phys.\ Rev.\ D {\bf 84}, 044052 (2011)
\href{http://xxx.lanl.gov/abs/1107.3173}{{\tt arXiv:1107.3173 [gr-qc]}}.

\bibitem{Guariento:2012ri} 
D.~C.~Guariento, M.~Fontanini, A.~M.~da Silva and E.~Abdalla,
{\it Realistic fluids as source for dynamically accreting black holes 
in a cosmological background},
Phys.\ Rev.\ D {\bf 86}, 124020 (2012)
\href{http://xxx.lanl.gov/abs/arXiv:1207.1086}{{\tt [arXiv:1207.1086 [gr-qc]]}}.

\bibitem{Rodrigues:2012xm} 
M.~G.~Rodrigues and A.~E.~Bernardini,
{\it Accretion of non-minimally coupled generalized Chaplygin gas into black holes},
Int.\ J.\ Mod.\ Phys.\ D {\bf 21}, 1250075 (2012)
\href{http://xxx.lanl.gov/abs/1208.1572}{{\tt arXiv:1208.1572 [gr-qc]}}.

\bibitem{Abdalla:2013ara}
E.~Abdalla, N.~Afshordi, M.~Fontanini, D.~C.~Guariento and E.~Papantonopoulos,
{\it Cosmological black holes from self-gravitating fields},
Phys.\ Rev.\ D {\bf 89} (2014) 104018
\href{http://xxx.lanl.gov/abs/1312.3682}{{\tt arXiv:1312.3682 [gr-qc]}}.

\bibitem{Babichev:2014lda} 
E.~O.~Babichev, V.~I.~Dokuchaev and Y.~N.~Eroshenko,
{\it Black holes in the presence of dark energy},
Phys.\ Usp.\  {\bf 56}, 1155 (2013)
[Usp.\ Fiz.\ Nauk {\bf 189}, no. 12, 1257 (2013)]
\href{http://xxx.lanl.gov/abs/1406.0841}{{\tt arXiv:1406.0841 [gr-qc]}}.

\bibitem{Afshordi:2014qaa} 
N.~Afshordi, M.~Fontanini and D.~C.~Guariento,
{\it Horndeski meets McVittie: A scalar field theory for accretion onto cosmological black holes},
Phys.\ Rev.\ D {\bf 90}, no. 8, 084012 (2014)
\href{http://xxx.lanl.gov/abs/1408.5538}{{\tt arXiv:1408.5538 [gr-qc]}}.

\bibitem{Davis:2016avf} 
A.~C.~Davis, R.~Gregory and R.~Jha,
{\it Black hole accretion discs and screened scalar hair},
JCAP {\bf 1610}, no. 10, 024 (2016)
\href{http://xxx.lanl.gov/abs/1607.08607}{{\tt arXiv:1607.08607 [gr-qc]}}.

\bibitem{Frolov:2017asg} 
A.~V.~Frolov, J.~T.~Gálvez Ghersi and A.~Zucca,
{\it Unscreening scalarons with a black hole},
Phys.\ Rev.\ D {\bf 95}, no. 10, 104041 (2017)
\href{http://xxx.lanl.gov/abs/1704.04114}{{\tt arXiv:1704.04114 [gr-qc]}}.

\bibitem{BCG}
P.~Bowcock, C.~Charmousis and R.~Gregory,
{\it General brane cosmologies and their global space-time structure},
Class.\ Quant.\ Grav.\  {\bf 17}, 4745 (2000)
\href{http://xxx.lanl.gov/abs/hep-th/0007177}{{\tt arXiv:hep-th/0007177}}.

\bibitem{Regge:1974zd} 
T.~Regge and C.~Teitelboim,
{\it Role of Surface Integrals in the Hamiltonian Formulation of General Relativity},
Annals Phys.\  {\bf 88}, 286 (1974).

\end{thebibliography}
\end{document}